\documentclass[journal,letterpaper,final]{IEEEtran}

\usepackage{graphicx}
\usepackage{color}
\usepackage{placeins}
\usepackage{float}
\usepackage{tabularx,colortbl}
\usepackage{amssymb}
\usepackage{amsthm}
\usepackage{cite}
\usepackage{amsmath}
\usepackage{subcaption}
\usepackage{algorithm}
\usepackage[noend]{algpseudocode}

\algrenewcommand\algorithmicindent{0.9em}%

\def\diag{\mathrm{diag}}

\def\Htran{\mbox{\tiny $\mathrm{H}$}}

\newlength{\figwidth}
\newcommand{\fracSum}[1]{{\underset{{#1}}{\sum}}}

\newcommand{\fracSumtwo}[2]{\overset{#2}{\underset{#1}{\sum}}}
\newcommand{\vect}[1]{\mathbf{#1}}

\newcommand{\br}[1]{{\left\{ #1 \right\}}}

\theoremstyle{plain}
\newtheorem{remark}{Remark}

\newtheorem{theorem}{Theorem}

\newtheorem{lemma}{Lemma}
\newtheorem{definition}{Definition}

\begin{document}

\title{{Adaptive Pilot Clustering in \\ Heterogeneous Massive MIMO Networks}}
\author{Rami~Mochaourab, \emph{Member, IEEE}, Emil~Bj\"ornson, \emph{Member, IEEE}, \\ and~Mats Bengtsson, \emph{Senior Member, IEEE}
\thanks{Part of this work has been presented at IEEE SPAWC, Stockholm, Sweden, June 28 - July 1, 2015 \cite{Mochaourab2015}. Rami Mochaourab and Mats Bengtsson are with ACCESS Linnaeus Centre, Signal Processing Department, School of Electrical Engineering, KTH Royal Institute of Technology, 100~44 Stockholm, Sweden. E-mail: \{rami.mochaourab, mats.bengtsson\}@ee.kth.se. Emil Bj\"ornson is with Department of Electrical Engineering (ISY), Link\"oping University, Sweden. E-mail: emil.bjornson@liu.se.}%
\thanks{R. Mochaourab and E. Bj\"ornson contributed equally to this work.}
\thanks{This research has received funding from the EU FP7 under ICT-619086 (MAMMOET) and was also supported by ELLIIT and CENIIT.}
\thanks{\copyright 2016 IEEE. Personal use of this material is permitted. Permission from IEEE must be obtained for all other uses, in any current or future media, including reprinting/republishing this material for advertising or promotional purposes, creating new
collective works, for resale or redistribution to servers or lists, or reuse of any copyrighted
component of this work in other works.}}

\IEEEoverridecommandlockouts

\maketitle

\begin{abstract}
We consider the uplink of a cellular massive MIMO network. Acquiring channel state information at the base stations (BSs) requires uplink pilot signaling. Since the number of orthogonal pilot sequences is limited by the channel coherence, pilot reuse across cells is necessary to achieve high spectral efficiency. However, finding efficient pilot reuse patterns is non-trivial especially in practical asymmetric BS deployments. We approach this problem using coalitional game theory. Each BS has a few unique pilots and can form coalitions with other BSs to gain access to more pilots. The BSs in a coalition thus benefit from serving more users in their cells, at the expense of higher pilot contamination and interference. Given that a cell's average spectral efficiency depends on the overall pilot reuse pattern, the suitable coalitional game model is in partition form. We develop a low-complexity distributed coalition formation based on individual stability. By incorporating a base station intercommunication budget constraint, we are able to control the overhead in message exchange between the base stations and ensure the algorithm's convergence to a solution of the game called individually stable coalition structure. Simulation results reveal fast algorithmic convergence and substantial performance gains over the baseline schemes with no pilot reuse, full pilot reuse, or random pilot reuse pattern.
\end{abstract}

\IEEEpeerreviewmaketitle


\section{Introduction}\label{sec:introduction}
The data traffic in cellular networks has increased exponentially for decades and this trend is expected to continue in the foreseeable future, spurred by new smart devices and innovative applications. The immense traffic growth has traditionally been handled by deploying more base stations (BSs) and allocating more frequencies for cellular communications. These approaches are less attractive in the future, since it is expensive to put a BS at every rooftop and because the spectral resources are scarce in the bands suitable for wide-area coverage (below 6 GHz). It is therefore important to also develop techniques that improve the spectral efficiency (bit/s/Hz/cell) in cellular networks, without requiring more BSs or additional frequency spectrum. The massive MIMO (multiple-input, multiple-output) concept was proposed in the seminal paper \cite{Marzetta2010a} as an attractive way to improve the spectral efficiencies of future networks by orders of magnitude.

In massive MIMO networks, the BSs are equipped with arrays with many active antenna elements (e.g., hundreds of small dipole antennas), which are processed coherently to improve the signal quality in both the uplink and the downlink\cite{Rusek2013a}. Massive MIMO is essentially a multi-user MIMO technology, thus it delivers high spectral efficiency by serving many user equipments (UEs) simultaneously. The performance per UE might not be higher than in contemporary networks, but the sum spectral efficiency per cell can be increased tremendously~\cite{Bjornson2016a}.

It is well known that multi-user MIMO systems require channel state information (CSI) at the BSs in order to separate the uplink signals sent in parallel by different UEs and to direct each downlink signal towards its intended receiver \cite{Gesbert2007a}. CSI can be acquired by sending predefined pilot sequences and estimate the channel responses from the received signals \cite{Bjornson2010a}. The pilot sequences are precious resources in cellular networks since accurate CSI estimation requires low interference in the pilot transmission phase (i.e., low so-called pilot contamination \cite{Jose2011b}). Contemporary networks have over-provision of pilot sequences---many more orthogonal pilots than active UEs per cell---thus the pilot contamination is essentially alleviated by selecting the pilots at random in every cell and switch the pilots regularly. In contrast, massive MIMO networks attempt to schedule as many users as possible to achieve a high sum spectral efficiency \cite{Bjornson2016a}. The number of pilot sequences then fundamentally limits the number of active UEs per cell.

{The early works on massive MIMO assumed that all pilot sequences were used in all cells, in which case one can only mitigate pilot contamination by exploiting spatial channel correlation as in \cite{Huh2012a,Yin2013a} or apply data-covariance-aided estimation methods \cite{Ngo2012a,Muller2014a}.\footnote{Time-shifting between data and pilot transmission has also been proposed to mitigate pilot contamination, but it has later been shown that also interfering data transmissions cause pilot contamination \cite[Remark 5]{Ngo2013a}.}} Recent works have shown that it is often beneficial to coordinate the pilot allocation with neighboring cells, for example, by having a non-universal pilot reuse to avoid pilot contamination from the first tier of interfering cells \cite{Yang2013a,Li2012a,Bjornson2016a}. This approach can make the impact of pilot contamination negligible for practical numbers of antennas, but at the cost of serving fewer UEs per cell---because only $1/3$, $1/4$, or $1/7$ of the pilot sequences are used in each cell. {If this is not enough, it can also be applied in conjunction with the data-covariance-aided methods proposed in \cite{Ngo2012a,Muller2014a} to further suppress interference.} This type of pilot allocation is conceptually simple in symmetric networks (e.g., one-dimensional cases as in \cite{Li2012a} or two-dimensional cases with hexagonal cells as in \cite{Saxena2015,Bjornson2016a}); one can cluster the cells by coloring them in a symmetric pattern and divide the pilot sequences so that only cells with the same color use the same subset of pilots. The clustering in practical asymmetric deployments, where every cell has a unique size and shape, is non-trivial and must be optimized for each particular deployment. 

Notice that pilot allocation problems are, in some respect, related to automated frequency assignment problems in cellular networks \cite{Hale1980}, which date back to the 1960's. These problems are known to be solvable using graph coloring algorithms, where a good assignment is characterized by low interference between cells having the same color (i.e., frequencies). It is not straightforward to apply frequency assignment algorithms for pilot allocation in massive MIMO, since these networks transmit data with universal frequency reuse. In contrast, an efficient pilot allocation mechanism for massive MIMO determines the number of pilots and scheduled UEs in each cell while taking the interference caused by all other cells into account \cite{Saxena2015}.

The purpose of this paper is to develop an algorithm for adaptive pilot clustering, which can be applied for decentralized optimization in cellular networks with arbitrary asymmetric cell geometries. To this end, we use tools from coalitional game theory~\cite{Osborne1994}. In our setting the set of players in the coalitional game correspond to the BSs, and a coalition between a set of players forms whenever they can take joint actions which lead to mutual benefits. In our setting, a set of BSs cooperate by sharing their pilot resources. Coalition formation games provide us with structured mechanisms to find the sets of cells which cooperate. Relying on rationality assumptions of the players, the mechanisms are naturally implementable in a distributed way. With such merits, coalitional game theory has found many applications in communication networks \cite{Saad2009,Saad2012,Guruacharya2013,Zhou2013,Mochaourab2014}. There are two types of coalitional game models: the characteristic form and the partition form \cite{Thrall1963}. In the characteristic form, the performance of a coalition assumes a predetermined behavior of the co-players not involved in the coalition. Coalitional games in partition form model the utility of each member of a coalition depending on the overall partition of the set of players, called the coalition structure. Since in our model, the performance of a coalition depends on the coalition structure, coalitional games in partition form are suitable in our context.

The solution of a coalitional game is a coalition structure which is stable according to a suitable stability model. In general, the stability is closely related to the method of deviation of the players; that is, the feasible ways to change from one coalition structure to another. Two stability models for coalition structures can be distinguished: group-based and individual-based stability. Group-based stability is satisfied if no set of players can jointly profit by changing the coalition structure and building a coalition together. Some applications of group-based stability solution concepts can be found in \cite{Saad2009,Guruacharya2013,Mochaourab2014}. In individual-based stability \cite{Dreze1980}, the change in the coalition structure occurs only when a single player leaves a coalition to join another. Thus, individual-based stability can be considered to be more restrictive than group-based stability in the deviation model and hence is generally of less complexity. Such stability concepts have been applied in \cite{Saad2012} in the context of cooperative channel sensing and access in cognitive radio and in \cite{Zhou2013} for cooperative precoding in the MIMO interference channel.

In this paper, we assume that each BS has a set of unique pilot sequences. A set of BSs can share their pilot sequences if they are in the same coalition, and consequently each BS in the coalition can schedule a larger number of UEs. However, the sharing of a BS's pilot resources creates pilot contamination effects within the coalition and might give a BS more pilots than it has UEs. Moreover, increasing the number of active UEs in the network increases the interference between the cells. In order to capture these effects, we first characterize the average sum SE of a cell depending on the underlying coalition structure. Based on the utility model, we propose a distributed coalition formation mechanism based on a model from \cite{Bogomolnaia2002}: each BS can leave its coalition and join another coalition if this strictly improves its average sum SE and does not reduce the average sum SE of the members of the coalition it joins. In order to control the complexity of the algorithm and guarantee its convergence to an individually stable coalition structure \cite{Bogomolnaia2002}, we define a base station intercommunication budget which limits the number of messages that can be sent from one base station to the other base stations during coalition formation. Simulation results reveal considerable performance gains using coalition formation over one-cell coalitions and universal pilot reuse.

\subsubsection*{Outline}
In Section~\ref{sec:system-model}, we describe the system model and derive the average sum spectral efficiency of a cell for a given coalition structure. The utility measures are utilized in Section~\ref{sec:coalition_formation} to formulate the coalitional game in partition form between the cells. Then, the coalition formation algorithm is specified and analyzed regarding stability and complexity. In addition, we provide a distributed implementation of the algorithm in the setting. In Section~\ref{sec:simulations}, we discuss the simulation results before we draw the conclusions in Section~\ref{sec:conclusion}.

\section{System Model \& Sum Spectral Efficiency}
\label{sec:system-model}

\begin{figure}[t]
  \centering
  \includegraphics[width=0.65\linewidth,clip]{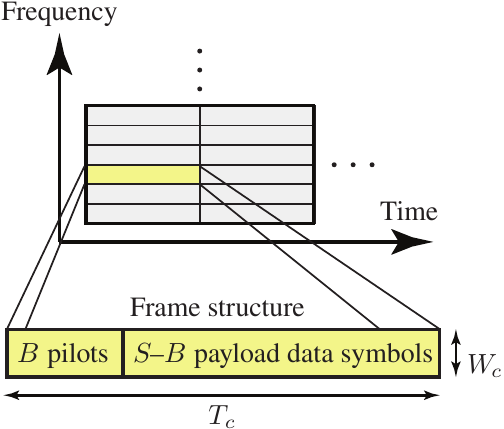}
  \caption{\label{fig:protocol} Frame structure in the uplink of a massive MIMO system, where $S = T_c W_c$ is the number of transmission symbols per frame.}
\end{figure}

We consider the uplink of a cellular massive MIMO network with $L$ cells, each assigned with an index in the set $\mathcal{L} = \{1, \ldots,L\}$. BS $j$ is equipped with an array of $M$ antennas and has a maximum of $K^\mathrm{max}_j$ connected single-antenna UEs. The data transmission is divided into frames of $T_c$ seconds and $W_c$ Hz, as illustrated in \figurename~\ref{fig:protocol}, which means that each frame contains $S = T_c W_c$ transmission symbols. The frame dimensions are matched to the coherence blocks of the channels so that the channel between each UE and each BS can be described by a constant channel response within a frame. In each uplink frame, $B$ symbols are allocated for pilot signaling and the remaining $S-B$ symbols are used for uplink payload data transmission.

The $B$ pilot symbols permit $B$ orthogonal pilot sequences; that is, only $B$ UEs in the network can transmit pilots without interfering with each other. {In this paper, we study how the $L$ cells should share these pilot sequences in order to maximize their sum spectral efficiency (SE), by balancing the number of active number of UEs and the degradation in channel estimation quality caused by having many UEs.} Since pilot contamination is mainly a problem in highly loaded networks, where many UEs in each cell are requesting data, this is the main focus of this paper. It is up to each BS to determine how many of its UEs that are active in each frame.

\subsection{Cell Cooperation Model}

We assume that each cell is given a fraction $B^\mathrm{cell} = \frac{B}{L}$ of unique pilot sequences,\footnote{This strategy is practically feasible if we consider the $L$ cells comprising a large cluster within a huge network in which the $B$ pilots are reused.} where $\frac{B}{L}$ for convenience is assumed to be an integer. BS $j$ can keep its $B^\mathrm{cell}$ pilots by itself and serve $B^\mathrm{cell}$ UEs without any pilot contamination. Alternatively, it can form a coalition with other cells to share the access to each others' pilots, and consequently serve more UEs.

We define the coalition concept as follows.

\begin{definition} \label{def:coalition-structure}
A \emph{coalition structure} $\mathcal{C}$ is a partition of $\mathcal{L}$, the grand coalition, into a set of disjoint coalitions $\{\mathcal{S}_1,\ldots,\mathcal{S}_N\}$ where $\bigcup_{n = 1}^{N} \mathcal{S}_n = \mathcal{L}$.
\end{definition}

\begin{figure}[t]
  \centering
  \includegraphics[width=\figwidth,clip]{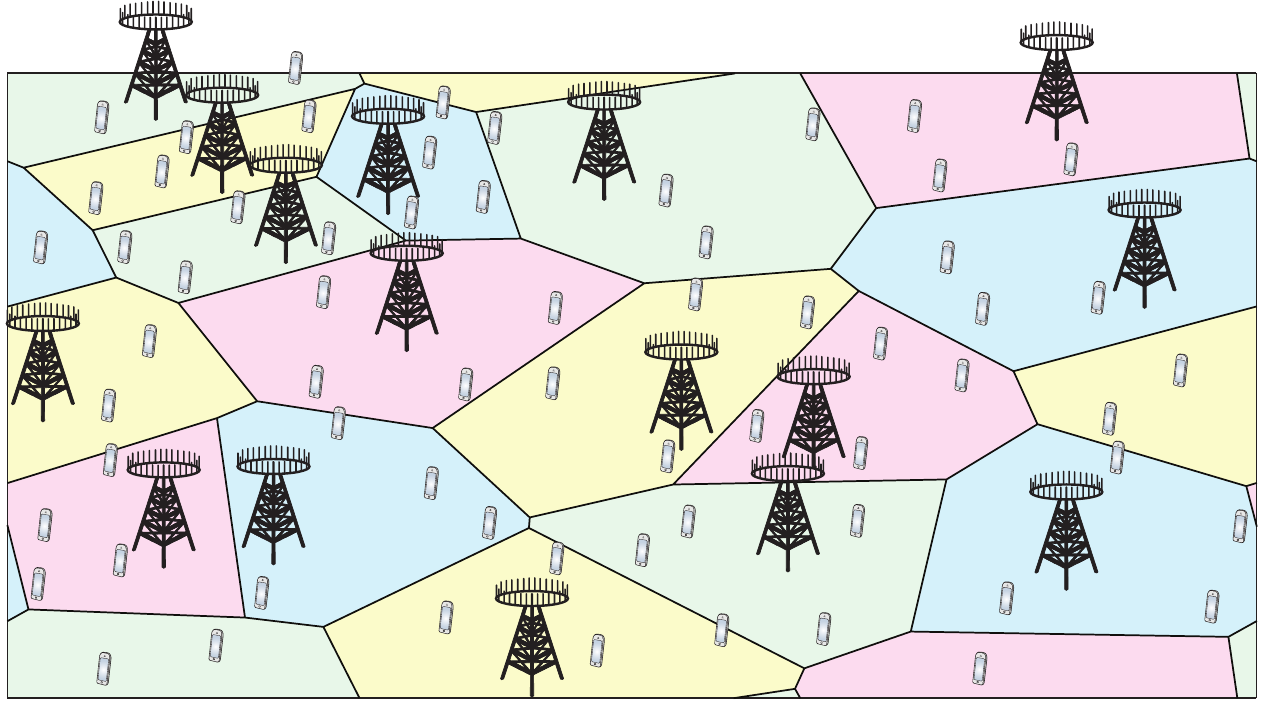}
  \caption{\label{fig:systemmodel}  Illustration of a cellular massive MIMO system with $L$ cells and wrap-around. Each cell contains a BS with $M$ antennas and a number of single-antenna UEs. The colors indicate coalitions of cells that use the same pilots.}
\end{figure}

For notational convenience, we let $\Phi_j (\mathcal{C})$ denote the coalition that BS $j$ belongs to for a given coalition structure $\mathcal{C}$. The members of the coalition $\Phi_j(\mathcal{C})$ have access to $| \Phi_j(\mathcal{C}) | B^\mathrm{cell}$ pilot sequences, where $| \cdot | \geq 1$ denotes the cardinality of a non-empty set (i.e., the number of set members). Then, the number of UEs that BS $j \in \Phi_j(\mathcal{C})$ can serve is 
\begin{equation}\label{eq:scheduling}
K_j(|\Phi_j(\mathcal{C})|) = \min \left\{ | \Phi_j(\mathcal{C}) | B^\mathrm{cell}, K_j^\mathrm{max} \right\}.
\end{equation}
However, the drawback is that cells in the same coalition contaminate each others pilot transmissions. \figurename~\ref{fig:systemmodel} gives an example of a cellular network with $L=16$ cells in a quadratic area. The cells are divided into four coalitions: green, yellow, red, and blue. Since each coalition has four members, each BS has access to ${4}B^\mathrm{cell}$ pilot sequences and BS $j$ serves exactly $K_j(4) = \min\{4 B^\mathrm{cell},K^\mathrm{max}_j\}$ UEs in each frame. Pilot contamination occurs only between cells with the same color. 

The coalition formation in this paper will determine a coalition structure $\mathcal{C}$ based on maximizing the SE in each cell. In the following, after describing basic assumptions in our uplink system model, we derive an expression for the average sum spectral efficiency of a cell for a given coalition structure$\mathcal{C}$. For notational convenience, we drop the dependency on $\mathcal{C}$ and write $\Phi_j$ and $K_j$ instead of $\Phi_j(\mathcal{C})$ and $K_j(|\Phi_j(\mathcal{C})|)$.

\subsection{Multi-Cell Channel Propagation}
\label{subsec:multi-cell-propagation}
The vast majority of prior works on massive MIMO assumes that each BS serves the same number of UEs (cf.~\cite{Marzetta2010a,Hoydis2013a,Ngo2013a,Bjornson2016a}). In contrast, the BSs in this paper may form coalitions of different sizes and thus serve unequal numbers of UEs. We therefore start from the beginning and provide the basic uplink system model for the problem at hand.

The UEs in a cell are picked at random from the coverage area and we will later consider the average performance over different UE distributions. In a certain frame, suppose that $\vect{z}_{lk} \in \mathbb{R}^2$ is the position of the $k$th UE in cell $l$. The channel response $\vect{h}_{jlk} \in \mathbb{C}^M$ between this UE and BS $j$ is modeled as Rayleigh fading:
\begin{equation} \label{eq:channel-distribution}
\vect{h}_{jlk} \sim \mathcal{CN}\Big(\vect{0},d_j(\vect{z}_{lk}) \vect{I}_M \Big),
\end{equation}
where $\vect{I}_M$ is the $M \times M$ identity matrix. The deterministic function $d_j(\vect{z})$ gives the variance of the channel attenuation from an arbitrary user position $\vect{z}$ to BS $j$. We assume that the value of $d_j(\vect{z}_{lk})$ is known at BS $j$ for all $l$ and $k$ (it is measured over frequency and tracked over time), while the exact UE positions are unknown. 

{The UEs use power control to counteract the average channel attenuation and achieve the same signal-to-noise ratio (SNR) to the serving BS irrespective of where the UE is. This is key to achieve uniform performance and avoid near-far issues in uplink multi-user MIMO. More precisely, we assume that a UE at position $\vect{z}_{jk}$ uses a transmit power of $\rho/ d_j(\vect{z}_{jk})$ per symbol, where $\rho$ is a design parameter and $d_j(\vect{z}_{jk})$ is the channel attenuation to the serving BS. The resulting average SNR at any antenna of the serving BS is $\rho/\sigma^2$, where $\sigma^2$ is the noise variance per symbol, and the average SINR also becomes the same for all UEs in a cell since the uplink interference that affect a UE is independent of its own position. The parameter $\rho$ is selected so that all UEs in the cells comply with their amplifier power constraints.}

Recall from \figurename~\ref{fig:protocol} that the first $B$ symbols of each frame are used for pilot transmission, which allows for $B$ orthogonal $B$-length pilot sequences. Each BS $j$ serves $K_j$ UEs and has access to $| \Phi_j | B^\mathrm{cell}$ pilot sequences, where \eqref{eq:scheduling} manifests that the number of UEs is always fewer or equal to the number of available sequences. To avoid cumbersome pilot coordination within the coalition, BS $j$ picks a subset of 
$K_j$ pilot sequences uniformly at random in each frame and distribute these among its UEs.
For some arbitrary UE $k$ in cell $j$ we let the random variable $\chi_{jklm}$ be 1 if UE $m$ in cell $l$ uses the same pilot sequence in a given frame and otherwise it is 0. The probability of $\chi_{jklm}=1$ is $\frac{1}{| \Phi_j | B^\mathrm{cell}}$ and the probability of $\chi_{jklm}=0$ is $1-\frac{1}{| \Phi_j | B^\mathrm{cell}}$.
Using this notation, the effective received pilot signal $\vect{y}_{jk}^{\mathrm{pilot}}\in \mathbb{C}^{M}$ at BS $j$ for its UE $k$ is 
\begin{multline} \label{eq:system-model-pilot}
\vect{y}_{jk}^{\mathrm{pilot}} = \sqrt{ \frac{\rho}{d_l(\vect{z}_{jk})} B } \vect{h}_{jjk} \\ + \sum_{l \in \Phi_j (\mathcal{C}) \setminus \{ j \} } \sum_{m=1}^{K_l} \chi_{jklm} \sqrt{ \frac{\rho}{d_l(\vect{z}_{lm})} B } \vect{h}_{jlm} + \boldsymbol{\eta}_{jk} \\ \mathrm{for} \,\, k = 1,\ldots,K_j,
\end{multline}
when BS $j$ has correlated the received signals with the pilot sequence used by its UE $k$ \cite{Hoydis2013a}.
The first term in \eqref{eq:system-model-pilot} is the desired signal and the last term $\boldsymbol{\eta}_{jk} \sim \mathcal{CN}(\vect{0},\sigma^2 \vect{I}_M)$ is the effective additive noise. The middle term is interference from UEs in cells of the coalition $\Phi_j$, while we stress that there is no interference from cells in other coalitions.

During uplink payload data transmission, all BSs are active and the received signal $\vect{y}_j \in \mathbb{C}^{M}$ at BS $j$ is 
\begin{equation} \label{eq:system-model}
\vect{y}_j^{\mathrm{data}} = \sum_{l = 1}^{L} \sum_{m=1}^{K_l} \sqrt{ \frac{\rho}{d_l(\vect{z}_{lm})} }  \vect{h}_{jlm} x_{lm} + \vect{n}_{j},
\end{equation}
where $x_{lm} \in \mathbb{C}$ is the data symbol transmitted by UE $k$ in cell $l$. This signal is normalized as $\mathbb{E}\{ | x_{lm} |^2 \} = 1$, while the corresponding UL transmit power is $ \frac{\rho}{d_l(\vect{z}_{lm})}$, as defined earlier. The additive receiver noise is modeled as $\vect{n}_{j} \sim \mathcal{CN}(\vect{0},\sigma^2 \vect{I}_M)$.

\subsection{Channel Estimation and Average Spectral Efficiency}

We will now compute closed-form achievable sum SE for each cell, which are later used for coalition formation in Section \ref{sec:coalition_formation}. As usual in massive MIMO, the BSs use coherent linear receive combining to detect the signals transmitted by each of the served UEs. This requires instantaneous CSI and we thus begin by stating the minimum mean-squared error (MMSE) of the channels from the received pilot signals in~\eqref{eq:system-model-pilot}.

\begin{lemma} \label{lemma:estimation}
The MMSE estimate of $\vect{h}_{jjk}$ at BS $j$ (for a given coalition structure $\mathcal{C}$ and given pilot allocations) is
\begin{equation} \label{eq:LMMSE-estimator}
  \hat{\vect{h}}_{jjk} =  \frac{ \sqrt{ \rho d_j(\vect{z}_{jk}) B }   }{\rho B +  \fracSum{\ell \in \Phi_j \setminus \{ j \} }  \,  \fracSumtwo{i=1}{K_\ell } \chi_{jk \ell i} \frac{\rho d_j(\vect{z}_{\ell i}) }{d_\ell(\vect{z}_{\ell i})} B + \sigma^2 } \vect{y}_{jk}^{\mathrm{pilot}}
\end{equation}
where $ \hat{\vect{h}}_{jjk} \sim \mathcal{CN}(\vect{0}, \delta_{jjk} \vect{I}_M )$ with the variance
\begin{equation}
\delta_{jjk} =  \frac{ \rho d_j(\vect{z}_{jk})  B   }{\rho B +  \fracSum{\ell \in \Phi_j \setminus \{ j \} }  \,  \fracSumtwo{i=1}{K_\ell } \chi_{jk \ell i} \frac{\rho d_j(\vect{z}_{\ell i}) }{d_\ell(\vect{z}_{\ell i})} B + \sigma^2 }.
\end{equation}
The estimation error $ \tilde{\vect{h}}_{jjk} = \vect{h}_{jjk} -\hat{\vect{h}}_{jjk} $ is independently distributed as 
\begin{equation}
\tilde{\vect{h}}_{jjk} \sim \mathcal{CN} \big(\vect{0}, (d_j(\vect{z}_{jk}) - \delta_{jjk}) \vect{I}_M \big).
\end{equation}

\end{lemma}
\begin{IEEEproof}
This lemma follows from applying standard results from \cite[Chapter 15.8]{Kay1993a} on MMSE estimation of Gaussian vectors in Gaussian colored noise.
\end{IEEEproof}

\begin{figure*}[t]
\begin{equation} \tag{14} \label{eq:SINR-value}
\mathrm{SINR}_{jk} =  \frac{ \frac{\rho}{ d_j(\vect{z}_{jk})} | \mathbb{E}_{\{\vect{h},\chi\}}\{ \vect{g}_{jk}^{\Htran} \vect{h}_{jjk} \} |^2 }{ \fracSum{l \in \mathcal{L}} \fracSumtwo{m=1}{K_l} 
\frac{\rho}{ d_l(\vect{z}_{lm})}
  \mathbb{E}_{\{\vect{h},\chi\}} \{ |\vect{g}_{jk}^{\Htran} \vect{h}_{jlm} |^2  \} - \frac{\rho}{ d_j(\vect{z}_{jk})} | \mathbb{E}_{\{\vect{h},\chi\}} \{ \vect{g}_{jk}^{\Htran} \vect{h}_{jjk} \} |^2  + \sigma^2  \mathbb{E}_{\{\vect{h},\chi\}} \{ \| \vect{g}_{jk} \|^2\}  }
\end{equation}
\noindent\rule{\textwidth}{0.4pt}
\begin{align} \tag{18} \label{eq:achievable-SINR-MR2}
I_{j}^{\mathrm{MRC}}(\mathcal{C}) = &\fracSum{l \in \Phi_j \setminus \{ j \}} \frac{K_l}{| \Phi_j | B^\mathrm{cell}} \left( \mu_{jl}^{(2)} + \frac{ \mu_{j l}^{(2)} - \left(\mu_{j l}^{(1)} \right)^2 }{M} \right) +
\left(  \sum_{\mathcal{S} \in \mathcal{C} } \sum_{l \in \mathcal{S}}
  \frac{K_l }{M}  \mu_{j l}^{(1)} +  \frac{\sigma^2}{M \rho}  \right)
 \left( 1 + \fracSum{\ell \in \Phi_j \setminus \{ j \}}  \frac{K_\ell}{| \Phi_j| B^\mathrm{cell}}  \mu_{j \ell}^{(1)}  + \frac{ \sigma^2 }{B \rho} \right)\\ \nonumber
I_{j}^{\mathrm{ZFC}}(\mathcal{C}) = & \fracSum{l \in \Phi_j \setminus \{ j \}} \frac{K_l}{| \Phi_j | B^\mathrm{cell}} \left( \mu_{jl}^{(2)}  + \frac{ \mu_{j l}^{(2)} - \left( K_l+1 \right) \left(\mu_{j l}^{(1)} \right)^2 }{M-K_j}  \right)   - \frac{K_j}{M-K_j} \\ \tag{19} \label{eq:achievable-SINR-ZF2} & + \left(\frac{ \sum\limits_{\mathcal{S} \in \mathcal{C} } \sum\limits_{l \in \mathcal{S}}
K_l \mu_{j l}^{(1)}  + \frac{\sigma^2}{\rho} }{M-K_j}\right)  \left( 1 + \fracSum{\ell \in \Phi_j \setminus \{ j \}}  \frac{K_\ell }{| \Phi_j | B^\mathrm{cell}}  \mu_{j \ell}^{(1)}  + \frac{ \sigma^2 }{B \rho} \right)
\end{align}
\noindent\rule{\textwidth}{0.4pt}
\end{figure*}

{Notice that Lemma \ref{lemma:estimation} gives the MMSE estimates of the UE channels within the serving cell, and characterizes the corresponding estimation errors. Each BS can also estimate channels to UEs in other cells of its coalition, for which we have the following lemma.}

\begin{lemma} \label{lemma:estimation-othercell}
If $\chi_{jk lm} = 1$ for some $l \in  \Phi_j$, then the MMSE estimate of $\vect{h}_{jlm}$ is
\begin{align}
\hat{\vect{h}}_{jlm} = \frac{d_j(\vect{z}_{lm})}{ \sqrt{ d_j(\vect{z}_{jk}) d_l(\vect{z}_{lm}) }}   \hat{\vect{h}}_{jjk},
\end{align}
where $ \hat{\vect{h}}_{jlm} \sim \mathcal{CN}(\vect{0}, \delta_{jlm} \vect{I}_M )$ has the variance
\begin{align}
\delta_{jlm} =   \frac{(  d_j(\vect{z}_{lm})  )^2}{ d_j(\vect{z}_{jk}) d_l(\vect{z}_{lm}) }   \delta_{jjk}  
\end{align}
and the independent estimation error is
\begin{align}
\tilde{\vect{h}}_{jlm} = \vect{h}_{jlm} -\hat{\vect{h}}_{jlm} \sim \mathcal{CN} \big(\vect{0}, (d_j(\vect{z}_{lm}) - \delta_{jlm}) \vect{I}_M \big).
\end{align}
\end{lemma}

This lemma shows the essence of pilot contamination, namely that $\hat{\vect{h}}_{jlm} $ and $\hat{\vect{h}}_{jjk} $ are equal up to a scaling factor when the corresponding UEs utilize the same pilot sequence. This important result is later used in the appendix when deriving SE expressions.

The linear detection at BS $j$ consists of assigning a combining vector $\vect{g}_{jk} \in \mathbb{C}^{M}$ to each of the $K_j$ UEs in the cell. By multiplying the received payload data signals in \eqref{eq:system-model} with these vectors, the effective scalar signal $\vect{g}_{jk}^{\Htran} \vect{y}_j^{\mathrm{data}} $ should amplify the intended signal $x_{jk}$ from the $k$th UE in the cell and/or suppress the interfering signals.

Let $\hat{\vect{H}}_{j} = [\hat{\vect{h}}_{jj1} \, \ldots \, \hat{\vect{h}}_{jj K_j}] \in \mathbb{C}^{M \times K_j}$ be a matrix with the estimated channels from Lemma  \ref{lemma:estimation} for the UEs in cell $j$. Two typical combining schemes are maximum ratio combining (MRC), which obtains the highest signal gain by setting
\begin{equation}
[\vect{g}_{j 1}^{\mathrm{MRC}} \, \ldots \,\, \vect{g}_{j K_j}^{\mathrm{MRC}}  ] =  \hat{\vect{H}}_{j} \vect{D}_{j} ,
\end{equation}
where $\vect{D}_{j} = \diag( M^{-1} \delta_{jj1}^{-1},\ldots, M^{-1}\delta_{jj K_j}^{-1})$ is a diagonal matrix\footnote{The normalization of $\hat{\vect{h}}_{jjk}$ by $M^{-1} \delta_{jjk}^{-1}$ in MRC makes the expected channel gain $\mathbb{E}\{ \vect{g}_{jk}^{\Htran} \vect{h}_{jjk} \} =1$ for both MRC and ZFC, and simplifies the derivations.}, and zero-forcing combining (ZFC) where the pseudo-inverse of $ \hat{\vect{H}}_{j}$ is used to suppress intra-cell interference:
\begin{equation}
[\vect{g}_{j 1}^{\mathrm{ZFC}} \, \ldots \,\, \vect{g}_{j K_j}^{\mathrm{ZFC}}  ] = \hat{\vect{H}}_{j} (  \hat{\vect{H}}_{j}^{\Htran}  \hat{\vect{H}}_{j}  )^{-1} .
\end{equation}

{In order to measure the data throughput in the cells, we use the ergodic capacity, which is the deterministic information rate that can be reliably communicated over a fading channel. The following lemma provides achievable sum SE expressions, applicable for any receive combining scheme including MRC and ZFC.}

\begin{lemma} \label{lemma:SE} 
Consider a given coalition structure $\mathcal{C}$, where cell $l$ serves $K_l$ UEs for all $l \in \mathcal{L}$. A lower bound on the average ergodic sum capacity achieved in cell $j$ is \begin{equation}
\mathrm{SE}_j =  \sum_{k=1}^{K_j}  \left( 1-\frac{B}{S} \right) \mathbb{E}_{\{\vect{z}\}} \left\{ \log_2(1+ \mathrm{SINR}_{jk}) \right\} ~\text{[bit/symbol]}
\end{equation}
which is a summation of the average ergodic SEs of the $K_j$ UEs in that cell. For given UE positions, the signal-to-interference-and-noise ratio (SINR) of the $k$th UE in cell $j$ is given in \eqref{eq:SINR-value} at the top of the page, where the expectation $\mathbb{E}_{\{\vect{h},\chi\}} \{ \cdot \}$ is with respect to the channel realizations and pilot allocations. The outer expectation $\mathbb{E}_{\{\vect{z}\}} \{ \cdot \}$ gives the average over different UE positions in the network.
\end{lemma}
\begin{IEEEproof}
Proved in the same way as \cite[Lemma 2]{Bjornson2016a}.
\end{IEEEproof}

{Note that Lemma \ref{lemma:SE} provides average ergodic sum SEs with respect to different UEs positions, different pilot allocations within the cells, small-scale fading variations, and CSI estimation errors.} It is a lower bound on the ergodic capacity, which is unknown for multi-cell scenarios with imperfect CSI. The pre-log factor has two parts: a summation over the number of active UEs in the cell $K_j$ given in \eqref{eq:scheduling} and the loss from the pilot signaling overhead $( 1 - \frac{B}{S} ) $. It is only the effective SINR, $\mathrm{SINR}_{jk}$, that depends on which receive combining scheme that is used in the network.

{Next, we use Lemma \ref{lemma:SE} to compute closed-form SE expressions for the MRC and ZFC schemes, when using the MMSE channel estimates obtained in Lemma \ref{lemma:estimation}. These expressions characterize the practically achievable data throughput per cell, which will later be used as utility functions.} As a preparation, we define the following propagation parameters:
\begin{align} \tag{15} \label{eq:mu-definition1}
\mu^{(1)}_{jl} &= \mathbb{E}_{\vect{z}_{lm}} \left\{  {d_j(\vect{z}_{lm}) }/{ d_l(\vect{z}_{lm})} \right\}, \\ \tag{16}\label{eq:mu-definition2}
\mu^{(2)}_{jl} &= \mathbb{E}_{\vect{z}_{lm}} \left\{ \big( {d_j(\vect{z}_{lm}) }/{ d_l(\vect{z}_{lm})} \big)^{2} \right\},
\end{align}
where the expectations are with respect to the arbitrary distribution of UE positions in cell $l$ and thus take the large-scale fading into account. The first one, $\mu^{(1)}_{jl}$, is the average ratio between the channel variance to BS $j$ and the channel variance to BS $l$, for a UE in cell $l$. The second one, $\mu^{(2)}_{jl} $, is the second-order moment of the same ratio. Hence, $\mu^{(2)}_{jl} - (\mu^{(1)}_{jl})^2$ is the variance of this ratio. Notice that $\mu^{(1)}_{jj} = \mu^{(2)}_{jj} =1$ when the two indices are the same, while the parameters are smaller than one when the indices are different. In general, we have $\mu^{(1)}_{jl} \neq \mu^{(1)}_{lj}$ while equality only holds for symmetric networks where the corresponding cells have the same shape. We use these propagation parameters to get the following result.

\begin{theorem} \label{theorem:sum-spectral-efficiency}
For a given coalition structure $\mathcal{C}$, a lower bound on the average ergodic sum capacity in cell $j$ is
\begin{equation} \tag{17} \label{eq:utility-functions}
U_j(\mathcal{C}) = \left( 1 - \frac{B}{S} \right) K_j \,\log_2 \left(  1 + \frac{1}{I_{j}^{\mathrm{scheme}}(\mathcal{C}) } \right)  \text{[bit/symbol]}
\end{equation}
where the interference term $I_{j}^{\mathrm{scheme}}$ with MRC and with ZFC (for $M>K_j$) are given  respectively in \eqref{eq:achievable-SINR-MR2} and \eqref{eq:achievable-SINR-ZF2} at the top of the page.

\end{theorem}
\begin{IEEEproof}
The proof is given in Appendix \ref{sec:proofth1}.
\end{IEEEproof}

\begin{figure}[t]
  \centering
  \includegraphics[width=\figwidth,clip]{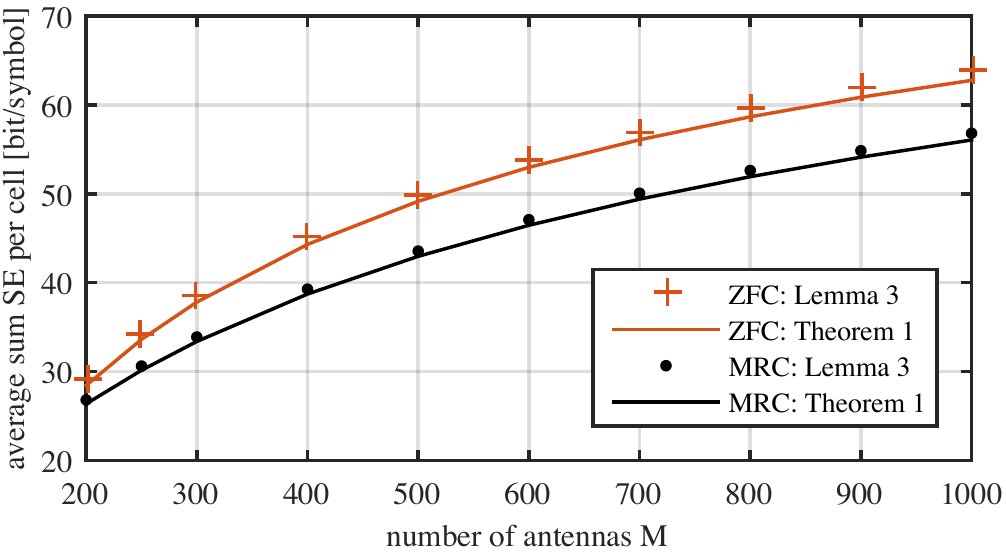}
  \caption{\label{fig:comparison_lb} Monte Carlo simulations for the average ergodic sum capacity in Lemma~\ref{lemma:SE} for comparison with the lower bound in Theorem~\ref{theorem:sum-spectral-efficiency}.} 
\end{figure}
{The closed-form lower bounds in Theorem \ref{theorem:sum-spectral-efficiency} are slightly more conservative than the non-closed-form bound in Lemma \ref{lemma:SE}. However, \figurename~\ref{fig:comparison_lb} shows that the difference is negligible when dealing with MRC and ZFC.}\footnote{This figure was generated using the same simulation setup as in Section \ref{sec:simulations} and \figurename~\ref{fig:SE_BS}, with $L=20$ cells, random coalitions, and varying number of BS antennas.} Note that maximizing the sum SE might lead to operating points with many active UEs and low SE per UE, but this is still beneficial for all UEs as compared to time-sharing where each UE is only active part of time but exhibit a higher SE when being active. Theorem~\ref{theorem:sum-spectral-efficiency} generalizes previous results in \cite{Bjornson2016a}, which only covered fixed pilot allocations and an equal number of UEs per cell.
Although the interference terms $I_{j}^{\mathrm{MRC}}(\mathcal{C}) $ and $I_{j}^{\mathrm{ZFC}}(\mathcal{C})$ have lengthy expressions, these are easy to implement and have intuitive interpretations. The first part of both expressions describes the pilot contamination and is only impacted by the cells that have formed a coalition with BS $j$. The second part describes the conventional inter-user interference (from all cells). MRC suppresses the impact of other signals and noise by amplifying the signal of interest using the full array gain of $M$, while ZFC only achieves an array gain of $M-K_j$ since BS $j$ sacrifices degrees of freedom for interference suppression within the cell. The interference suppression results in the extra negative term on the first row in \eqref{eq:achievable-SINR-ZF2}, and ZFC is preferable over MRC whenever the reduced interference is more substantial than the loss in array gain. Which of the schemes that provide the highest performance varies depending on the SNR, the coalition design, and how strong the interference is between the cells.

The average sum SE $U_j(\mathcal{C})$ in \eqref{eq:utility-functions} for cell $j$ should preferably be as large as possible. This is the utility function that we assign to BS $j$ in the remainder of this work. There are thus $L$ different utilities and their values depend on the selection of combining scheme (e.g., MRC or ZFC) and on the coalition structure $\mathcal{C}$. {The average sum SE $U_j(\mathcal{C})$ is based on the ergodic capacities of UEs at different locations, which connect and disconnect to the network over time. This is a good utility if the data package transmitted by each UE spans many fading realization (over time and frequency) or when $M$ is large so that fading averages out due the channel hardening. We envision a coalition formation mechanism that makes decisions based on the average throughput over time intervals of at least a second.\footnote{Shorter coherence times and more intermittent user activity enable shorter time intervals since there is more randomness.} Since $U_j(\mathcal{C})$ depends on the number of UEs available in the cells, the coalition structure $\mathcal{C}$ should be updated when the number of UEs changes significantly. Small-scale variations in the number of users occur at the millisecond level (due to bursty traffic), but the important large-scale variations occur over the hours of the day \cite{Auer2011b}. The network designer can select how often it is worth to re-optimize the coalition structure.}

From the structure of the pilot contamination terms, it is preferable for a BS to form coalitions with cells that are far away, but this intuition is hard to transform to any simple algorithm for coalition formation, except for completely symmetric cellular networks as in \cite{Bjornson2016a}. For general asymmetric networks the system designer can, in principle, traverse all possible coalition structures, but unfortunately the number of possibilities equals the $L$th Bell number, which has a faster growth than exponential with $L$. Consequently, finding a globally optimal pilot assignment is hard. In the next section, we therefore formulate the design problem as a coalitional game and provide an efficient decentralized algorithm to find stable coalition structures.

\begin{remark}[Uplink-downlink duality]
The average ergodic SE in Theorem \ref{theorem:sum-spectral-efficiency} is for the uplink, but can also be used to describe the downlink. There is a property called uplink-downlink duality that, basically, says that the same sum SE can be achieved in both directions---using the same total transmit power, but with different power allocation over the UEs. The classic duality concept was established in \cite{Viswanath2003a} and \cite{Boche2002a} for perfect CSI, and it was generalized to massive MIMO in \cite{Bjornson2016a}. As a consequence, network optimization (e.g., coalition formation) based on the uplink formulas in Theorem \ref{theorem:sum-spectral-efficiency} optimizes also the downlink.
\end{remark}
\setcounter{equation}{19}

\section{Coalitional Game}\label{sec:coalition_formation}
We analyze in this section cooperation between the BSs using coalitional games. The strategies of the BSs are directly related to the coalition they are members of; that is, they share their pilots with the cells in their coalition. {Since the average SE of each cell, given in Theorem~\ref{theorem:sum-spectral-efficiency}, depends on the coalition structure (Definition~\ref{def:coalition-structure}), we need to study the coalitional game in partition form~\cite{Thrall1963}, which we formulate by 
\begin{equation}
\langle \mathcal{L}, \tilde{\vect{U}} \rangle.
\end{equation}
Here, the set of players corresponds to the set of BSs $\mathcal{L}$. Let $\mathcal{P}$ be the set of all partitions of $\mathcal{L}$. The partition function $\tilde{\vect{U}} : \mathcal{P} \rightarrow \mathbb{R}^{L}$ assigns a payoff to each player for each partition in $\mathcal{P}$ and will be formulated shortly in \eqref{eq:restricted_utility}.}

\begin{figure*}[t]
  \centering
  \includegraphics[width=0.75\textwidth,clip]{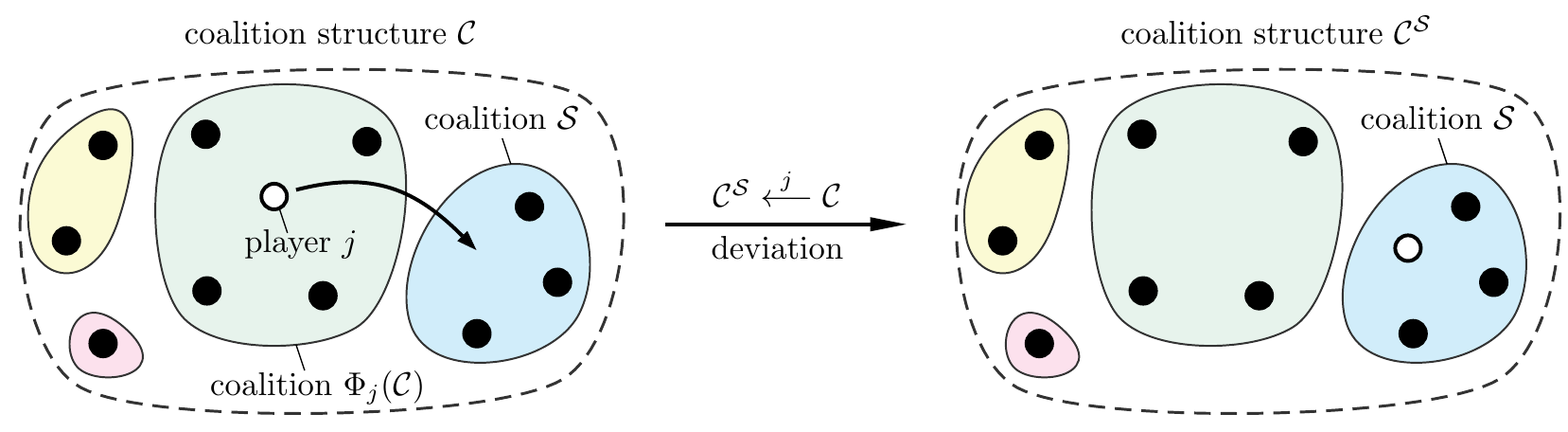}
  \caption{\label{fig:deviation} Illustration of the deviation model.}
\end{figure*}

When considering cooperation between the BSs (players), we assume that they are able to communicate with each other and exchange specific application-type messages. This is possible in cellular networks due to the existence of backhaul links connecting the BSs \cite{Gesbert2010a}. However, there is a cost involved in communication between the BSs which is quantified in terms of the number of data packets exchanged among them. In order to limit this overhead, we introduce a \emph{base station intercommunication budget} $q_j\in\mathbb{N}$ for each BS $j$ in $\mathcal{L}$, which limits the total number of data packets sent from BS $j$ to the other BSs during coalition formation. With this constraint, the nontransferable utility of a player $j$ is formulated as
\begin{equation} \label{eq:restricted_utility}
\tilde{U}_j(\mathcal{C},\eta_j) = \begin{cases}
   U_j(\mathcal{C}) &\text{if } \eta_j \leq q_j,\\
   0       &\text{otherwise},\\
  \end{cases}
\end{equation}
where $U_j(\mathcal{C})$ is the average sum SE of cell $j$ given in~Theorem~\ref{theorem:sum-spectral-efficiency}, and $\eta_j \in\mathbb{N}$ represents the number of data packets which player $j$ has already sent to the other BSs. The utility model in \eqref{eq:restricted_utility} gives a player $j$ zero utility if it has exhausted its BS intercommunication budget $q_j$. Note that, besides suitably controlling the messaging overhead between the base stations, the intercommunication budget will be used to ensure convergence of the coalition formation algorithm (Algorithm~\ref{alg:coalition0}) where the parameters $\eta_j$ are updated.\footnote{Another way to ensure convergence of the considered coalition formation algorithm is by associating with each player a history set which is utilized to prevent the player from joining a coalition it has been a member of before, as in e.g. \cite{Saad2009b,Saad2012, Zhou2013,Guazzone2014}.}

From Theorem~\ref{theorem:sum-spectral-efficiency}, the utility of cell $j$ depends on which members are in its coalition $\Phi_j(\mathcal{C})$ through the pilot contamination term as well as the interference term determined by the coalitions formed outside $\Phi_j(\mathcal{C})$. Therefore, so-called \emph{externalities} exist \cite{Yi1997}. More specifically, our game relates to the category of coalitional games with negative externalities in which the merging of coalitions reduces the utility of the members of all coalitions not involved in the merging. This occurs due to the increased number of scheduled UEs and thereby increasing the interference.

In considering a coalitional game, we adopt the game theoretic assumptions which imply that each player's behavior follows the maximization of its utility function in~\eqref{eq:restricted_utility} based on the discovery of profitable opportunities \cite{Osborne1994}. Such behavior is important for the distributed implementation of our solution concept which we specify and discuss next. We stress that our solution is not limited to the performance measures in Theorem~\ref{theorem:sum-spectral-efficiency} but can be utilized in conjunction with any other utility function (e.g., utility functions that take other types of channel fading into account).

\subsection{Coalition Formation}
Coalition formation describes the dynamics which lead to stable coalition structures. We use a coalition formation model from \cite{Bogomolnaia2002} in which a single player is allowed to leave its coalition and join another only if it is profitable for the player and all members of the coalition it wants to join. Such a coalition formation model has been used, e.g. in \cite{Saad2012} in the context of cognitive radio settings.

Three elements are needed to describe our coalition formation game \cite{Apt2006,Saad2009a,Mochaourab2014}: 1) a deviation model; 2) a comparison relation which indicates whether a deviation is acceptable; and 3) a stability concept for coalition structures.

\begin{definition}[Deviation]\label{def:deviation}
A cell $j\in \mathcal{L}$ leaves its current coalition $\Phi_j(\mathcal{C})$ to {join} coalition $\mathcal{S} \in \mathcal{C} \cup \{\emptyset\}$. In doing so, the coalition structure $\mathcal{C}$ changes to $\mathcal{C}^{\mathcal{S}}$. We capture this change in the coalition structure by the notation $\mathcal{C}^{\mathcal{S}} \overset{j}{\longleftarrow} \mathcal{C}$.
\end{definition}

An illustration of the deviation model is given in \figurename~\ref{fig:deviation}. {Observe that a deviation entitles a player to search for alternatives within the current coalition structure. Given a coalition structure $\mathcal{C}$, the number of searches by a player $j$ is upper bounded by 
\begin{equation}\label{eq:complexity_dev}
D_j(\mathcal{C}) = \begin{cases} |\mathcal{C}| & \text{if } |\Phi_j(\mathcal{C})| > 1\\ 
|\mathcal{C}| - 1 & \mbox{otherwise}. \end{cases}
\end{equation}
\noindent The two cases in \eqref{eq:complexity_dev} differ by the possibility whether player $j$ can join the empty set or not, where the latter case is not relevant when player $j$ is in a singleton coalition. The worst case deviation complexity corresponds to the coalition structure in which all players are in singleton coalitions. Then, $D_j(\{\{1\},\ldots,\{L\}\}) = L-1$ which is linear in the number of players.}

According to the individual stability concept in \cite{Bogomolnaia2002}, a deviation is admissible if a player can strictly improve its performance by leaving its current coalition to join another coalition ensuring that the members of the coalition it joins do not reduce their utility. 
\begin{definition}[Admissible deviation]\label{def:admissable}
A deviation $\mathcal{C}^{\mathcal{S}} \overset{j}{\longleftarrow} \mathcal{C}$ is admissible if 
\begin{multline}
\tilde{U}_j(\mathcal{C}^{\mathcal{S}},\eta_j) > \tilde{U}_j(\mathcal{C},\eta_j) \text{ and } \tilde{U}_k(\mathcal{C}^{\mathcal{S}},\eta_k) \geq \tilde{U}_k(\mathcal{C},\eta_k),\\ \text{ for all } k \in \mathcal{S}.
\end{multline}
\end{definition}
Such a deviation requirement is suitable for our setting due to the fact that each cell exclusively owns a set of pilots and any BS that wants to join a coalition by sharing its pilots with the coalition members must first ask their permission. 

Following the players' rationality assumption, admissible deviations according to Definition~\ref{def:deviation} will be pursued by the players. Accordingly, we utilize the following stability concept for coalition structures \cite{Bogomolnaia2002}.

\begin{definition}[Individual stability]\label{def:Individual_stability}
A coalition structure $\mathcal{C}$ is individually stable if there exists no $j\in\mathcal{L}$ and coalition $\mathcal{S}$ such that $\mathcal{C}^{\mathcal{S}} \overset{j}{\longleftarrow} \mathcal{C}$ is admissible.
\end{definition}

\begin{algorithm}[t]
\caption{\label{alg:coalition_formation} Coalition formation algorithm.}
\begin{algorithmic}[1]
\Statex \textbf{Initialize}: $t = 0$, coalition structure $\mathcal{C}_{0}$; 
\Repeat
	\State Find a player $j \in \mathcal{L}$ and a coalition $\mathcal{S} \in \mathcal{C}_{t}$;
	\If{deviation $\mathcal{C}^{\mathcal{S}}_{t} \overset{j}{\longleftarrow} \mathcal{C}_{t}$ is admissible}
		\State Update coalition structure $\mathcal{C}_{t+1} = \mathcal{C}^{\mathcal{S}}_{t}$;
		\State Increment coalition index: $t = t+1$;
	\EndIf
\Until No deviation is admissible
\end{algorithmic} 
\end{algorithm}%

A generic coalition formation algorithm which terminates at a individually stable coalition structure is stated in Algorithm~\ref{alg:coalition_formation}. The algorithm is initialized with an arbitrary coalition structure $\mathcal{C}_{0}$. In each iteration, a player $j$ and a coalition $\mathcal{S}$ are selected to check whether their deviation is admissible according to Definition~\ref{def:admissable}. If so, the coalition structure changes according to the deviation. Since Algorithm~\ref{alg:coalition_formation} iterates over all BS deviation opportunities, an individually stable coalition structure is reached whenever the algorithm converges. The distributed implementation of Algorithm~\ref{alg:coalition_formation}, which we provide next, is guaranteed to converge since we impose the intercommunication budget constraint for each player as in \eqref{eq:restricted_utility}.

\subsection{Distributed Algorithm}

\begin{algorithm}[t]
\caption{\label{alg:coalition0} Implementation of coalition formation.}
\begin{algorithmic}[1]
\Statex \textbf{Initialize}: $t = 0$, $\mathcal{C}_{0} = \br{\br{1},\ldots,\br{K}}$, $\eta_j = 0, j \in \mathcal{L}$;
\Repeat
	\ForAll{BSs $j \in \mathcal{L}$}
	\State find acceptable coalitions $$\mathcal{D}_j = \{ \mathcal{S} \in \mathcal{C}_{t} \mid \tilde{U}_j(\mathcal{C}^{\mathcal{S}}_{t}, \eta_j) > \tilde{U}_j(\mathcal{C}_{t},\eta_j),\mathcal{C}^{\mathcal{S}}_{t} \overset{j}{\longleftarrow} \mathcal{C}_{t}\}$$
			\ForAll{$\mathcal{S} \in \mathcal{D}_j$ selected in random order} 
			\If{$\eta_j \leq q_j$}
				\State Ask members of $\mathcal{S}$ for permission to join;
				\State Increment the BS intercommunication cost;
				\ForAll{BSs $k \in \mathcal{S}$}
					\If{$\tilde{U}_k(\mathcal{C}^{\mathcal{S}}_{t}, \eta_k) \geq \tilde{U}_k(\mathcal{C}_{t},\eta_k)$}
						\State BS $k$ accepts;
						\State Increment the BS intercommunication cost;
					\EndIf
				\EndFor
			\EndIf

			\If{all BSs $k \in \mathcal{S}$ accept BS $j$}
				\State BS $j$ leaves $\Phi_j(\mathcal{C}_t)$ and joins $\mathcal{S}$;
				\State Update coalition structure $\mathcal{C}_{t+1} = \mathcal{C}^{\mathcal{S}}_{t}$;
				\State Increment the coalition index: $t = t + 1$;
				\State Inform all BSs about new coalition structure;
				\State Increment the BS intercommunication cost;
			\EndIf
		\EndFor
	\EndFor
\Until No cell deviates
\end{algorithmic} 
\end{algorithm}%

In Algorithm~\ref{alg:coalition0}, we provide an implementation of Algorithm~\ref{alg:coalition_formation}. { We initialize the coalition structure in $\mathcal{C}_{0}$, which can be the singleton coalitions (corresponding to no pilot reuse) or any other coalition structure.} A BS $j$ is selected at random to check if a deviation is profitable. Based on the local knowledge of the current coalition structure $\mathcal{C}_t$ and the propagation parameters, BS $j$ can calculate its utility in~\eqref{eq:restricted_utility} if it joins other coalitions in $\mathcal{D}_j \subseteq \mathcal{C}_t$. Note that $\mathcal{D}_j$ includes only the coalitions in which BS $j$ would strictly profit by joining. 

BS $j$ selects coalitions $\mathcal{S} \in \mathcal{D}_j$ at random\footnote{Random selection of a coalition $\mathcal{S}$ in $\mathcal{D}_j$ is reasonable given the uncertainty that $\mathcal{S}$ would accept player $j$ to join its coalition.} in Line~4. If $\eta_j$ satisfies the budget constraint (Line~5), BS $j$ sends a message, included in a single data packet, to the members of coalition $\mathcal{S}$ indicating that it wants to join their coalition (Line~6) and increments its BS intercommunication cost in Line~7. Each BS $k\in\mathcal{S}$ can calculate its utility locally for the case that BS $j$ enters the coalition. If the utility of BS $k\in\mathcal{S}$ does not decrease (Line~9), then BS $k$ accepts BS $j$ (Line~10) by sending its decision in a message to BS $j$. Here, BS $k$ increments its BS intercommunication cost in Line~11. Otherwise, BS $k$ does not reply to BS $j$'s request. If all BSs in coalition $\mathcal{S}$ accept BS $j$ (Line~12), meaning that the deviation is admissible (Definition~\ref{def:admissable}), then BS $j$ leaves its coalition and joins $\mathcal{S}$. The coalition structure is updated in Line~14 and all BSs are informed of the new coalition structure through a message from BS $j$. Algorithm~\ref{alg:coalition0} terminates when no deviations take place anymore.

\begin{figure*}[t!]
    \centering
    \begin{subfigure}[t]{0.49\textwidth}
        \centering
        \includegraphics[width=\textwidth,clip]{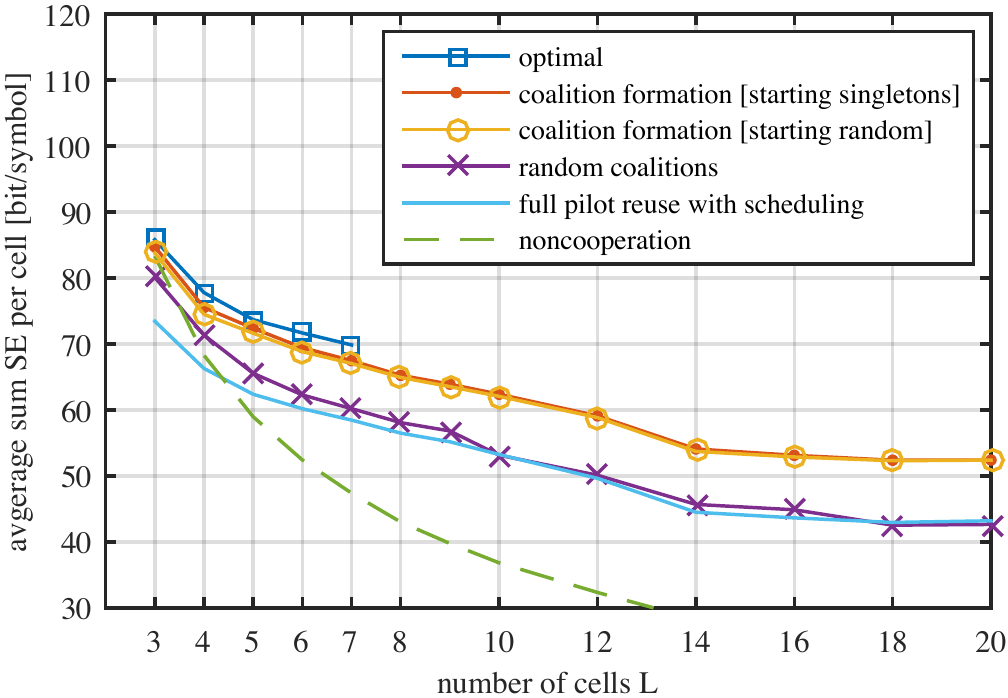}
  \caption{\label{fig:SE_BS_MRC} MRC}
    \end{subfigure}%
    ~\hspace{0.01\textwidth}
    \begin{subfigure}[t]{0.49\textwidth}
        \centering
        \includegraphics[width=\textwidth,clip]{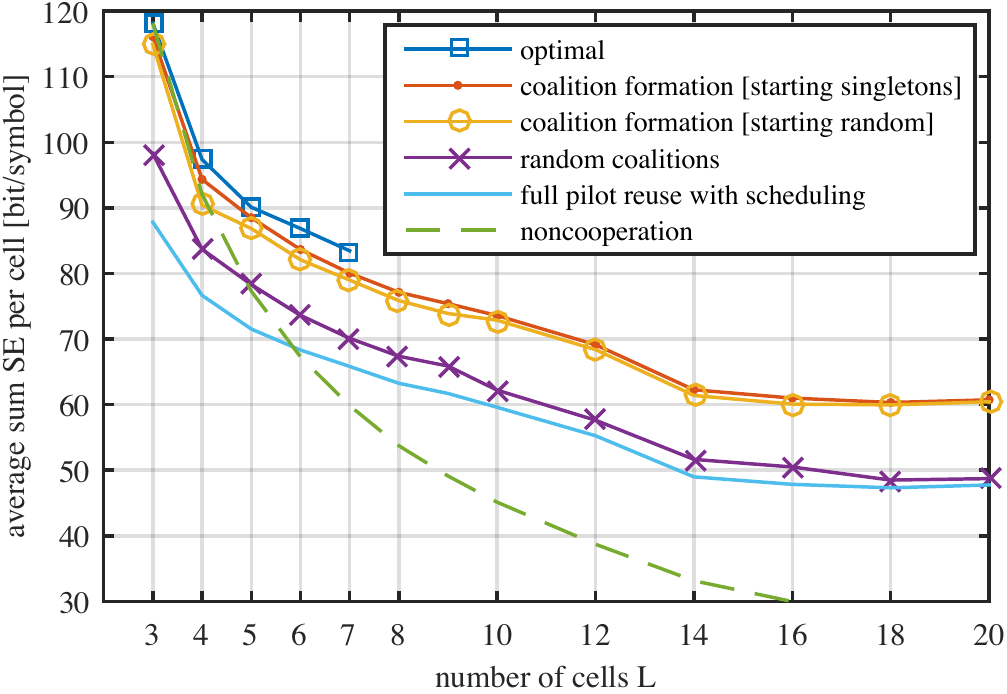}
  \caption{\label{fig:SE_BS_ZFC} ZFC}
    \end{subfigure}
    \caption{\label{fig:SE_BS} Average SE per cell for different number of cells $L$ with $M = 500$ and $K^\mathrm{max} \geq 200$.}
\end{figure*}

\begin{figure*}[t!]
    \centering
    \begin{subfigure}[t]{0.49\textwidth}
      \centering
      \includegraphics[width=\textwidth,clip]{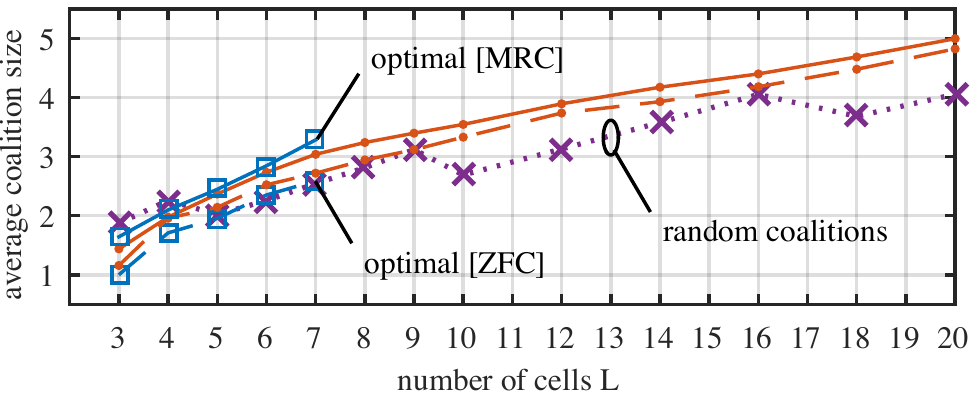}
      \caption{\label{fig:SIZE_BS} Average coalition sizes}
    \end{subfigure}%
    ~\hspace{0.01\textwidth}
    \begin{subfigure}[t]{0.49\textwidth}
      \centering
      \includegraphics[width=\textwidth,clip]{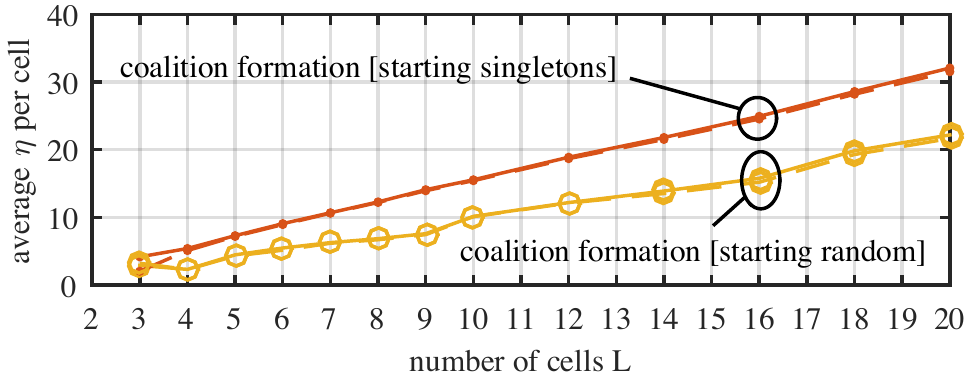}
      \caption{\label{fig:ETA_BS} Average number of messages sent per base station during coalition formation}
\end{subfigure}%
    \caption{\label{fig:SIZE_ETA_BS} Coalition sizes and number of searches per cell associated with the curves in \figurename~\ref{fig:SE_BS} for different number of cells $L$ with $M = 500$ and $K^\mathrm{max} \geq 200$. The solid (dashed) curves correspond to MRC (ZFC).}
\end{figure*}
\section{Simulations}\label{sec:simulations}%
In this section, we illustrate the coalition formation by simulations. {We consider frames with $S = 400$ symbols (e.g., $T_c \!=\!4$ ms and $W_c\!=\!100$ kHz) and a pathloss exponent of $3$. The SNR per receive antenna at the BS is SNR = $\frac{\rho}{\sigma^2}$ $=5$ dB, which is achieved for every UE by virtue of the power control policy described in Section~\ref{subsec:multi-cell-propagation}.} 

{We assume that each BS owns $B^\textrm{cell} = \lfloor \frac{\alpha S}{L}\rfloor$ unique pilot sequences where $\alpha$ determines the fraction of the frame used for pilot signaling. In the simulations, we set $\alpha = 0.5$ except for the plot in Section~\ref{sec:dependency_B}.} The available number of UEs in each cell is chosen to be the same: $K_j^\mathrm{max} = K^\mathrm{max}$ for all $j \in \mathcal{L}$. When considering different number of cells, we ensure the same BS density of $25$ BSs/km$^2$ by appropriately choosing the region area the cells are deployed in.

For the coalition formation algorithm, we do not include the intercommunication budget constraint for the base stations, except in Section~\ref{sec:dependency_q}. Although there is no guarantee for convergence of the coalition formation algorithm without this budget restriction, we witnessed convergence in all simulation instances. {We initialize the coalition formation algorithm using two different coalition structures, to evaluate the impact of the initialization. One initial coalition structure is singleton coalitions which corresponds to no pilot reuse. The other initial coalition structure is generated randomly with an average coalition size of $\lceil \sqrt{L} \rceil$, which is roughly the average coalition size that the coalition formation algorithm achieves in the simulations. 

Throughout, we compare pilot clustering according to coalition formation to three schemes: 
\begin{itemize}
\item The first scheme uses no pilot reuse which is called noncooperation. 
\item The second scheme corresponds to random coalition structures with average coalition size of $\lceil \sqrt{L} \rceil$. Having this scheme, which has similar average coalition sizes as coalition formation, highlights the importance of the selection of the members of each coalition.
\item The third scheme corresponds to using all available pilots by all the cells (full pilot reuse) but with user scheduling dictated by the coalition formation algorithm initialized with singleton coalitions. Including this scheme, with similar scheduling as coalition formation, emphasizes the importance of the pilot reuse patterns.
\end{itemize}}

We obtain the average performance using $2 \times 10^3$ uniformly random BS deployments with uniform user distributions in each cell and a wrap-around topology, as exemplified in Fig.~\ref{fig:systemmodel}. 

\subsection{Number of Cells}

In \figurename~\ref{fig:SE_BS_MRC} and \figurename~\ref{fig:SE_BS_ZFC}, the average sum SE per cell obtained in Theorem~\ref{theorem:sum-spectral-efficiency} with the MRC and ZFC schemes are plotted, respectively, for different number of BSs. Note here that we ensure the same BS density as discussed at the beginning of this section. The number of UEs per cell is chosen sufficiently large such that $K^\mathrm{max} \geq L B^\mathrm{cell}$. The optimal solution, which includes an exhaustive search over all possible coalition structures (corresponding to the $L$th Bell number), can be calculated for up to $L=7$ cells.

Random coalition formation whose average coalition size is $\lceil \sqrt{L} \rceil$, has similar performance as the full pilot reuse scheme. Both schemes are outperformed by the coalition formation algorithms, and it can be observed that the gains from coalition formation slightly increase with the size of the network. Initializing coalition formation with singletons gives slightly better performance than starting in random coalition structures. 

The average coalition sizes achieved by coalition formation starting in singletons are shown in \figurename~\ref{fig:SIZE_BS}. The results for coalition formation starting in random coalitions are similar and thus omitted. It can be noticed that the average coalition sizes are relatively small compared to the size of the network and scales roughly as $\lceil \sqrt{L} \rceil$. This result has influenced the choice of the average coalition sizes for the random scheme and random initialization. In comparison to MRC, ZFC has slightly smaller average coalition sizes. The reason for this is ZFC favors scheduling smaller number of users in the cells compared to MRC since ZFC expends a larger amount of the available spatial degrees of freedom for interference nulling.

The average number of messages sent per BS during coalition formation, corresponding to the plots in~\figurename~\ref{fig:SE_BS}, is shown in \figurename~\ref{fig:ETA_BS} and is observed to be very small compared to the size of the network. Clearly, initializing the algorithm in random coalitions leads to faster convergence and less message exchange between the base stations than starting in singletons.

\subsection{Number of BS antennas}\label{sec:dependency_M}
\begin{figure}[t]
  \centering
  \includegraphics[width=\figwidth,clip]{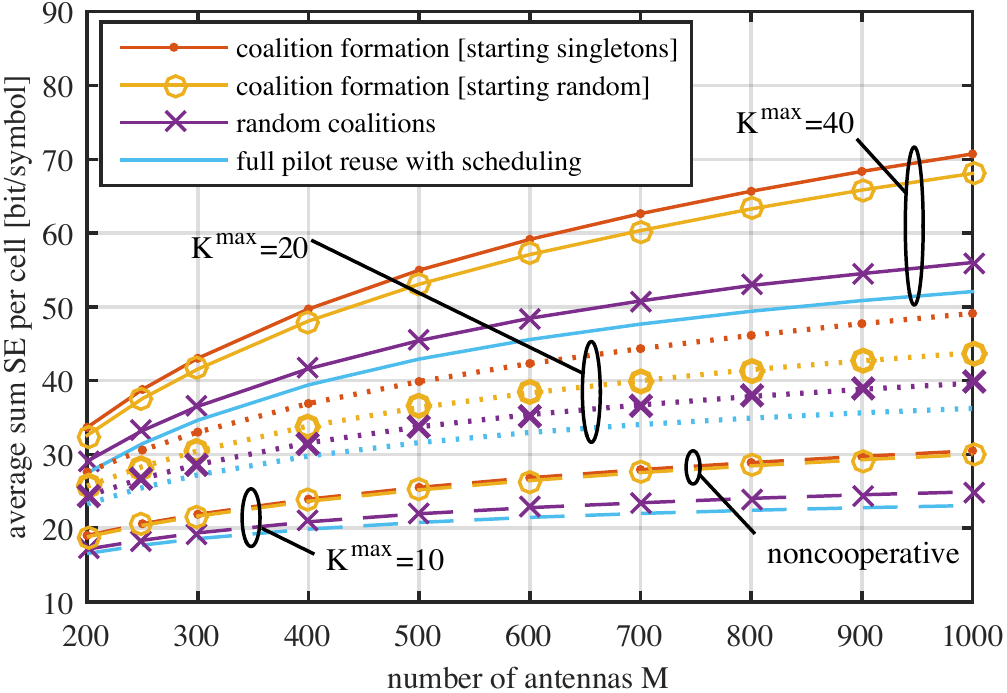}
  \caption{\label{fig:SE_M_ZFC_L20} Average sum spectral efficiency per cell using ZFC for different number of antennas at the BSs $M$. The number of cells is $L=20$.}
\end{figure}

In \figurename~\ref{fig:SE_M_ZFC_L20}, the average sum SE per cell with ZFC is plotted for different number of BS antennas $M$ and different number of available UEs $K^\mathrm{max}$ in each cell. The qualitative performance achieved using MRC is comparable to that of ZFC and is hence omitted. Coalition formation generally outperforms the other schemes and the gains in coalition formation increase for larger number of antennas.

For $K^\mathrm{max}=10$, noncooperation is optimal since each cell can schedule all available UEs with the pilots it possesses, given $B^\mathrm{cell} = \lfloor \frac{S}{2 L}\rfloor = 10$. In this case, both coalition formation with different initializations terminate in singletons. For larger $K^\mathrm{max}$, initializing coalition formation in singletons has better performance than starting in random coalitions as can be seen for $K^\mathrm{max}> 10$. This effect is studied in more detail in~Section~\ref{sec:dependency_K} when we study the dependency of the performance on $K^\mathrm{max}$. 

The average coalition sizes associated with~\figurename~\ref{fig:SE_M_ZFC_L20} do not show dependence on the used number of antennas $M$. We reveal these depending on $K^\mathrm{max}$ also in the next section for fixed $M = 500$.

\subsection{Available Users in Each Cell}\label{sec:dependency_K}
\begin{figure}[t!]
    \centering
    \begin{subfigure}[t]{\linewidth}
      \centering
      \includegraphics[width=\figwidth,clip]{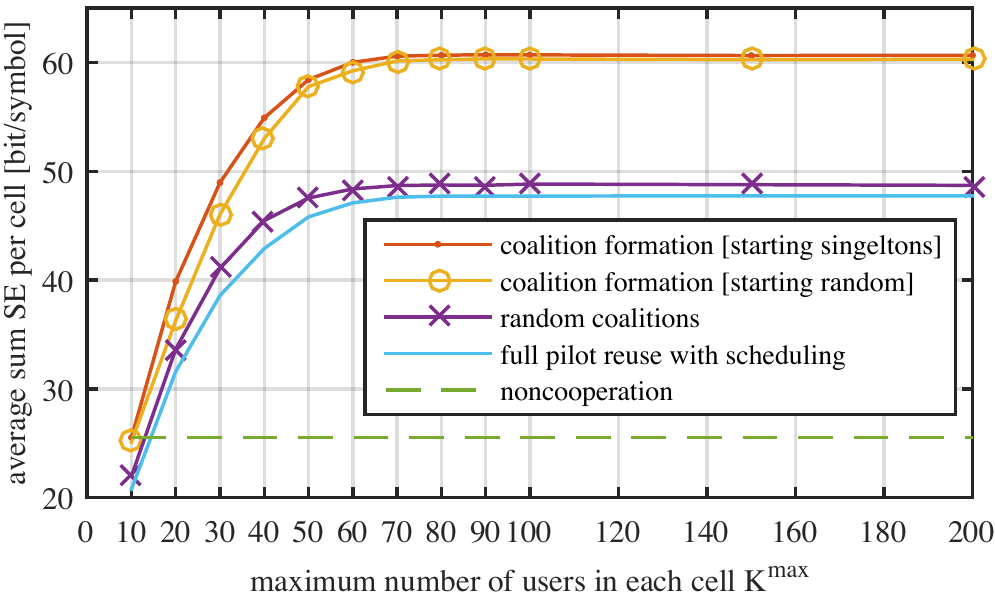}
      \caption{\label{fig:SE_K_ZFC} Average sum SE per cell}
    \end{subfigure}%
    \vspace{1em}
    \begin{subfigure}[t]{\linewidth}
      \centering
      \includegraphics[width=\figwidth,clip]{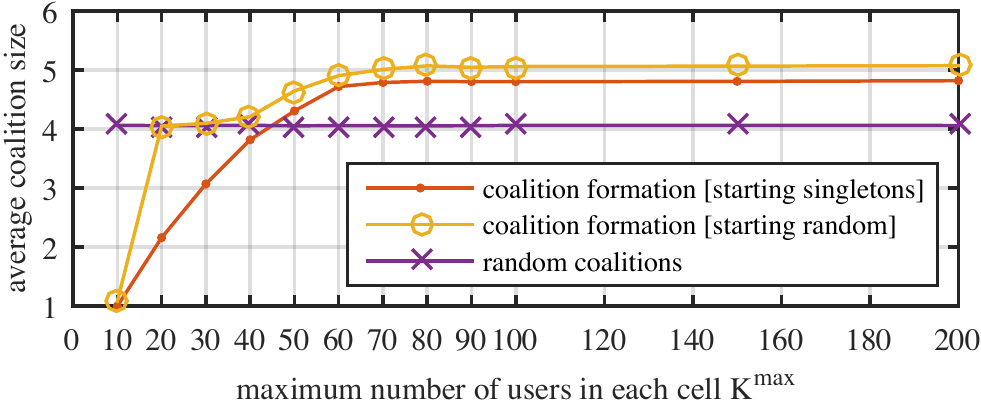}
      \caption{\label{fig:SIZE_K_ZFC} Average coalition sizes}
\end{subfigure}%
    \caption{\label{fig:SE_SIZE_K} Performance using ZFC for different number of available users in each cell. The number of cells is $L=20$ and the number of antennas at each BS is $M=500$.}
\end{figure}
\figurename~\ref{fig:SE_K_ZFC} shows how the gains in coalition formation increase with the number of users in each cell $K^\mathrm{max}$. Observe that the number of scheduled users with full pilot reuse is the same as coalition formation starting in singletons. Also, the random coalitions scheme and coalition formation which is initialized in random coalitions have almost similar number of scheduled users in the cells which can be seen from the average coalition sizes in~\figurename~\ref{fig:SIZE_K_ZFC}.

For $K^\mathrm{max} = 10$, noncooperation is optimal since each cell can schedule all its users with the available pilots. Starting in singletons, no deviations occur during coalition formation while starting in random coalitions, the BSs are able to leave their initial coalitions to form singleton coalitions. For $K^\mathrm{max} = 20$, and having $B^\mathrm{cell} =10$, it is sufficient to build coalitions of sizes of two to schedule all UEs in the cells. Coalition formation starting random has average coalition sizes of about four which is not efficient as is seen in~\figurename~\ref{fig:SE_M_ZFC_L20} and \figurename~\ref{fig:SE_K_ZFC}. Therefore, coalition formation starting with singleton coalitions is favored over random coalition structure initialization.

For coalition formation starting in singletons, the average coalition size in~\figurename~\ref{fig:SIZE_K_ZFC} is slightly larger than two although coalitions with two members are sufficient to schedule all users. Here, the BSs profit from excess pilots in order to reduce the pilot contamination in the coalition. Recall, that we assume random allocation of the pilots among the BSs in the same coalition when the number of UEs in the cell is less than the available pilots. 

For large values of $K^\mathrm{max}$, larger coalitions enable scheduling a larger number of users. Both coalition formation algorithms have similar performance and converge to similar average coalition sizes as is shown in~\figurename~\ref{fig:SIZE_K_ZFC}. Compared to the other schemes, the gains with coalition formation increase with $K^\mathrm{max}$.

\subsection{Base Station Intercommunication Budget}\label{sec:dependency_q}
\begin{figure}[t]
  \centering
  \includegraphics[width=\figwidth,clip]{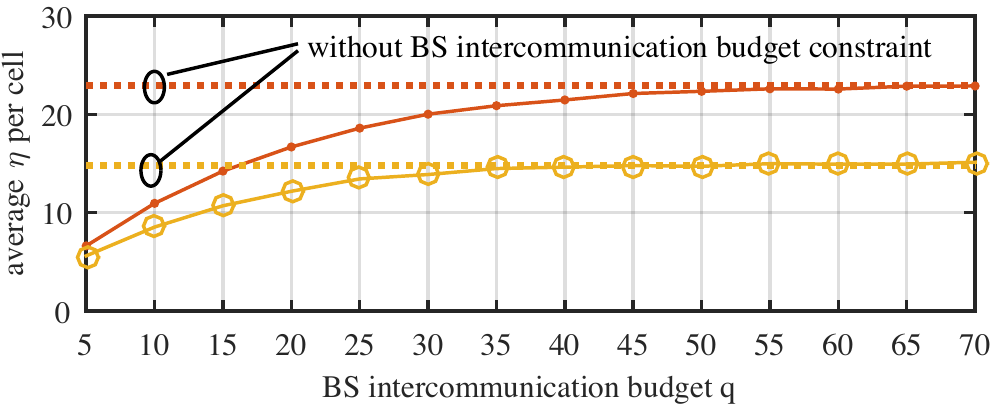}
  \caption{\label{fig:ETA_q_ZFC}Average number of messages sent per BS during coalition formation, depending on the BS intercommunication budgets $q = q_1,\ldots,q_L$ for $L=20$ BSs and $M=500$ antennas.}
\end{figure}
The complexity of coalition formation is reflected by the average number of messages sent from each BS, illustrated in~\figurename~\ref{fig:ETA_q_ZFC}. The number of messages, $\eta_j$, that a BS $j$ sends during coalition formation is incremented in Algorithm~\ref{alg:coalition0}. As seen in~\figurename~\ref{fig:ETA_q_ZFC}, the curves saturate at a relatively low value and meet the corresponding limits (dashed curves) which correspond to coalition formation without the budget restriction.

\subsection{Number of Pilots}\label{sec:dependency_B}

\begin{figure}[t]
  \centering
  \includegraphics[width=\figwidth,clip]{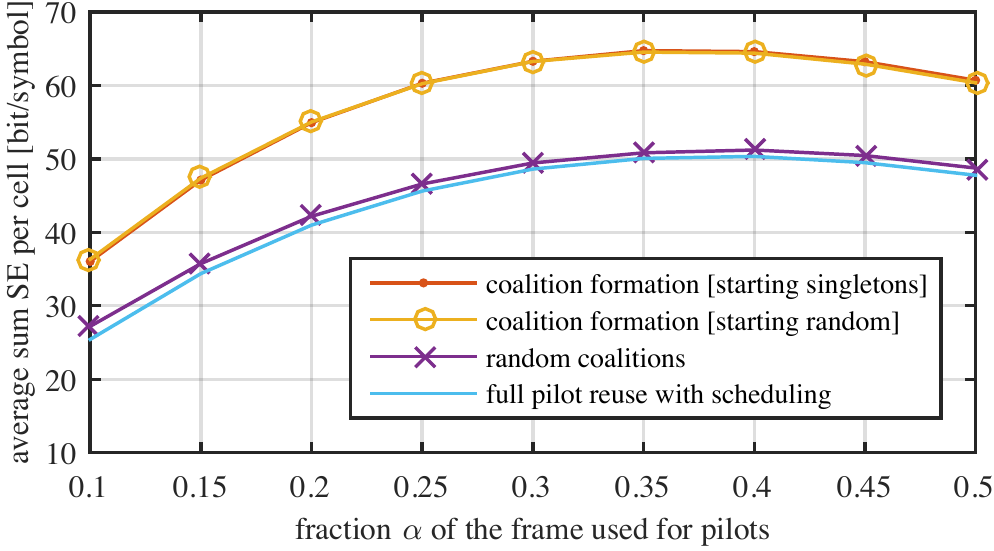}
  \caption{\label{fig:SEvsB_ZFC_BS20}Average sum SE per cell using ZFC depending on the fraction of the frame duration $S = 400$ used for pilots. The number of cells is $L=20$ and the number of antennas is $M=500$.}
\end{figure}

In~\figurename~\ref{fig:SEvsB_ZFC_BS20}, we show the effects of changing $\alpha$ which determines the fraction of the frame used for pilot signaling. The optimal choice of $\alpha$ is strictly less than $0.5$ and will generally depend on the number of antennas $M$. The outcome from coalition formation gives high performance adapting to different choices of $\alpha$.

\section{Conclusion}\label{sec:conclusion}

A distributed coalition formation algorithm is proposed in this paper for pilot allocation in the uplink of cellular massive MIMO networks with arbitrary geometry. Each BS is assumed to possess a few unique pilots which can be shared with other BSs in a coalition. The sharing of pilot resources permits scheduling more UE in the cooperating cells, however at the cost of increased pilot contamination effects and interference. We address the problem of finding the sets of cooperating BSs using coalitional games in partition form, by taking the number of pilots, the number of available users in each cell, the CSI quality, the receive combining scheme (MRC or ZFC), and the interference into account. The proposed coalition formation algorithm is based on an individual stability solution concept whose distributed implementation is shown to require low communication overhead between the BSs. Hence, it can both be used for offline cell planning and for online coalition formation that adapts the system to the small-scale user load variations that occur at the millisecond level due to bursty traffic or to the natural large-scale traffic variations over the day. Performance gains are achieved over baseline pilot reuse schemes for different network sizes and number of antennas at the BSs.

While the numerical results are based on the closed-form utility functions that were derived in this paper, we stress that the proposed coalition formation algorithm can be applied for many other utilities functions as well.
\appendices

\section{Proof of Theorem \ref{theorem:sum-spectral-efficiency}}\label{sec:proofth1}

In the case of MRC, the expectations in \eqref{eq:SINR-value} can be computed directly as
\begin{equation*}
\begin{split}
\mathbb{E}_{\{\vect{h},\chi\}}\{ \vect{g}_{jk}^{\Htran} \vect{h}_{jjk} \} & = \mathbb{E}_{\{\vect{h},\chi\}}\left\{ \frac{1}{M \delta_{jjk}} \hat{\vect{h}}_{jjk}^{\Htran} \hat{\vect{h}}_{jjk} \right\} \\ & = \mathbb{E}_{\{\chi\}}\left\{ \frac{M \delta_{jjk}}{M \delta_{jjk}} \right\}= 1
\end{split}
\end{equation*}
and in \eqref{eq:exp_gain}, at the top of next page, 
\begin{figure*}[th]
\begin{multline}\label{eq:exp_gain}
\mathbb{E}_{\{\vect{h},\chi\}} \{ \| \vect{g}_{jk} \|^2\} = \mathbb{E}_{\{\vect{h},\chi\}} \left\{  \frac{\| \hat{\vect{h}}_{jjk} \|^2}{ M^2 \delta_{jjk}^2} \right\} = \mathbb{E}_{\{\chi\}} \left\{  \frac{M \delta_{jjk} }{ M^2 \delta_{jjk}^2} \right\} = \frac{1}{M} \mathbb{E}_{\{\chi\}} \left\{ \frac{1}{\delta_{jjk}} \right\} \\ 
= \frac{1}{M}  \mathbb{E}_{\{\chi\}} \left\{ 
\frac{\rho B +  \fracSum{\ell \in \Phi_j \setminus \{ j \} }  \,  \fracSumtwo{i=1}{K_\ell } \chi_{jk \ell i} \frac{\rho d_j(\vect{z}_{\ell i}) }{d_\ell(\vect{z}_{\ell i})} B + \sigma^2 }{ \rho d_j(\vect{z}_{jk}) B   } \right\} 
= \frac{1}{M} \frac{1}{d_j(\vect{z}_{jk})} \underbrace{\left( 1 + \sum_{\ell \in \Phi_j (\mathcal{C}) \setminus \{ j \}} \sum_{i=1}^{K_\ell} \frac{1}{| \Phi_j | B^\mathrm{cell}}  \frac{ d_j(\vect{z}_{\ell i}) }{d_\ell(\vect{z}_{\ell i})}   + \frac{ \sigma^2 }{B \rho} \right)}_{=A_j},
\end{multline}
\noindent\rule{\textwidth}{0.4pt}
\end{figure*}
where the notation $A_j $ was introduced for brevity. In this computation we used the fact that $\mathbb{E}\{ \chi_{jk l m} \} = \frac{1}{| \Phi_j | B^\mathrm{cell}}$.

Moreover, for $l \not \in  \Phi_j$, or for $l=j$ with $k\neq m$, we have
\begin{equation*}
\begin{split}
 \mathbb{E}_{\{\vect{h},\chi\}} \{ |\vect{g}_{jk}^{\Htran} \vect{h}_{jlm} |^2  \} & = d_j(\vect{z}_{lm}) \mathbb{E}_{\{\vect{h},\chi\}} \{ \|\vect{g}_{jk} \|^2 \} = \frac{A_j}{M} \frac{d_j(\vect{z}_{lm})}{d_j(\vect{z}_{jk})} 
 \end{split}
\end{equation*}
since the MRC vector is independent of the channels in other coalitions and the channels of other UEs in the same cell.
For $l \in  \Phi_j$ we can perform the calculations in \eqref{eq:exp_gain2} at the top of next page, 
\begin{figure*}[th]
\begin{equation} \label{eq:exp_gain2}
\begin{split}
\mathbb{E}_{\{\vect{h},\chi\}} \{ |\vect{g}_{jk}^{\Htran} \vect{h}_{jlm} |^2  \} & =
 \mathbb{E}_{\{\chi\}} \left\{
  (1-\chi_{jklm}) \frac{d_j(\vect{z}_{lm})}{M^2 \delta_{jjk}^2} \mathbb{E}_{\{\vect{h}\}} \{ \| \hat{\vect{h}}_{jjk} \|^2 \}  \right\}  \\
  & \quad +  \mathbb{E}_{\{\chi\}} \left\{
  \frac{\chi_{jklm}}{M^2 \delta_{jjk}^2} \left(  ( d_j(\vect{z}_{lm})  - \delta_{jlm}) \mathbb{E}_{\{\vect{h}\}} \{\| \hat{\vect{h}}_{jjk} \|^2 \}  +   \frac{(  d_j(\vect{z}_{lm})  )^2}{ d_j(\vect{z}_{jk}) d_l(\vect{z}_{lm}) } 
  \mathbb{E}_{\{\vect{h}\}} \{ \| \hat{\vect{h}}_{jjk} \|^4 \} 
   \right)
 \right\} \\
 &=  \mathbb{E}_{\{\chi\}} \left\{
  (1-\chi_{jklm}) \frac{d_j(\vect{z}_{lm})}{M \delta_{jjk}} + 
  \chi_{jklm} \left(  \frac{( d_j(\vect{z}_{lm})  - \delta_{jlm})}{M \delta_{jjk}} +   \frac{(  d_j(\vect{z}_{lm})  )^2}{ d_j(\vect{z}_{jk}) d_l(\vect{z}_{lm}) } \frac{\delta_{jjk}^2 (M + M^2)}{M^2 \delta_{jjk}^2}
   \right)
 \right\}  \\
  &=  \mathbb{E}_{\{\chi\}} \left\{
  \frac{d_j(\vect{z}_{lm})}{M \delta_{jjk}} + 
  \chi_{jklm} \frac{(  d_j(\vect{z}_{lm})  )^2}{ d_j(\vect{z}_{jk}) d_l(\vect{z}_{lm}) }
 \right\} = \frac{d_j(\vect{z}_{lm})}{M d_j(\vect{z}_{jk})} A_j + \frac{1}{| \Phi_j | B^\mathrm{cell}} \frac{(  d_j(\vect{z}_{lm})  )^2}{ d_j(\vect{z}_{jk}) d_l(\vect{z}_{lm}) }
 \end{split}
\end{equation}
\noindent\rule{\textwidth}{0.4pt}
\end{figure*}
where the first equality follows from separating the two cases $\chi_{jklm}=0$ and $\chi_{jklm}=1$, where $\vect{g}_{jk}$ and $\vect{h}_{jlm}$ are independent in the first case and parallel in the second case; see Lemma \ref{lemma:estimation-othercell}. The second equality in \eqref{eq:exp_gain2} follows from computing the expectations with respect to the channel fading, where $\mathbb{E}_{\{\vect{h}\}} \{ \| \hat{\vect{h}}_{jjk} \|^4 \} $ is computed using \cite[Lemma 2]{Bjornson2015b}. The third inequality follows from some simple algebra and the last equality from the fact that $\mathbb{E}\{ \chi_{jk l m} \} = \frac{1}{| \Phi_j | B^\mathrm{cell}}$. 

By plugging these expectations into \eqref{eq:SINR-value} and dividing all terms with $\frac{\rho}{d_j(\vect{z}_{jk})}$, we obtain 
\begin{align} \label{eq:SINR-MRC}
&\mathrm{SINR}_{jk} = \\ \notag
& \frac{ 1 }{ 
\fracSum{l \in \Phi_j \setminus \{ j \}} \fracSumtwo{m=1}{K_l} 
 \frac{ (d_j(\vect{z}_{lm}) )^2}{  (d_l(\vect{z}_{lm}) )^2}  \frac{1}{| \Phi_j | B^\mathrm{cell}}  +
  \fracSum{l \in \mathcal{L} } \fracSumtwo{m=1}{K_l} 
  \frac{1}{M} \frac{d_j(\vect{z}_{lm})}{d_l(\vect{z}_{lm})}  A_j +  \frac{\sigma^2}{\rho} \frac{A_j }{M} } .
\end{align}

The expression in \eqref{eq:achievable-SINR-MR2} is obtained by computing an achievable lower bound $$\mathbb{E}_{\{\vect{z}\}} \{\log_2(1+ \frac{1}{f(\{\vect{z}\})}) \} \geq \log_2(1+ \frac{1}{\mathbb{E}_{\{\vect{z}\}} \{ f(\{\vect{z}\}) \} }) $$ where the expectation with respect to user positions are moved to the denominator of the SINR in \eqref{eq:SINR-MRC}. These exceptions are computed as follows:
\begin{equation} \label{eq:position-averaging1}
\mathbb{E}_{\{\vect{z}\}} \left\{ \fracSumtwo{m=1}{K_l}  \frac{ (d_j(\vect{z}_{lm}) )^2}{  (d_l(\vect{z}_{lm}) )^2}  \frac{1}{| \Phi_j | B^\mathrm{cell}}  \right\} = \frac{K_l}{| \Phi_j | B^\mathrm{cell}} \mu_{jl}^{(2)}
\end{equation}
\begin{equation} \label{eq:position-averaging2}
\begin{split}
\mathbb{E}_{\{\vect{z}\}} \left\{ A_j \right\} & = \mathbb{E}_{\{\vect{z}\}} \left\{ 1 + \sum_{\ell \in \Phi_j \setminus \{ j \}} \sum_{i=1}^{K_\ell} \frac{1}{| \Phi_j | B^\mathrm{cell}}  \frac{ d_j(\vect{z}_{\ell i}) }{d_\ell(\vect{z}_{\ell i})} + \frac{ \sigma^2 }{B \rho}  \right\} \\ & = 1 + \sum_{\ell \in \Phi_j \setminus \{ j \}}  \frac{K_\ell}{| \Phi_j | B^\mathrm{cell}}  \mu_{j \ell}^{(1)}  + \frac{ \sigma^2 }{B \rho}
\end{split}
\end{equation}
\begin{multline} \label{eq:position-averaging3}
\mathbb{E}_{\{\vect{z}\}} \left\{  \fracSum{l \in \mathcal{L} } \fracSumtwo{m=1}{K_l} 
  \frac{1}{M} \frac{d_j(\vect{z}_{lm})}{d_l(\vect{z}_{lm})}  A_j  \right\} = \\ \fracSum{l \in \mathcal{L} } 
  \frac{K_l}{M}  \mu_{j l}^{(1)}   \left( 1 + \sum_{\ell \in \Phi_j \setminus \{ j \}}  \frac{K_\ell}{| \Phi_j | B^\mathrm{cell}}  \mu_{j \ell}^{(1)}  + \frac{ \sigma^2 }{B \rho} \right) \\ +
   \sum_{l \in \Phi_j \setminus \{ j \}}  \frac{K_l}{| \Phi_j | B^\mathrm{cell}}  \frac{\left( \mu_{j l}^{(2)} - (\mu_{j l}^{(1)})^2 \right)}{M}
\end{multline}

Similarly, the expectations in \eqref{eq:SINR-value} can be computed for ZFC as
\begin{equation}
\mathbb{E}_{\{\vect{h},\chi\}}\{ \vect{g}_{jk}^{\Htran} \vect{h}_{jjk} \} = \mathbb{E}_{\{\vect{h},\chi\}}\{ \vect{g}_{jk}^{\Htran} \hat{\vect{h}}_{jjk} \} = 1
\end{equation}
and
\begin{multline}
\mathbb{E}_{\{\vect{h},\chi\}} \{ \| \vect{g}_{jk} \|^2\} = \mathbb{E}_{\{\vect{h},\chi\}} \{ [ (  \hat{\vect{H}}_{j}^{\Htran}  \hat{\vect{H}}_{j}  )^{-1} ]_{kk} \} \\ = \frac{1}{M-K_j}  \mathbb{E}_{\{\chi\}} \left\{ \frac{1}{\delta_{jjk}} \right\}
 = \frac{1}{M-K_j} \frac{A_j}{d_j(\vect{z}_{jk})}
\end{multline}
which follow from the zero-forcing definition and by utilizing well-known properties of Wishart matrices (see e.g., \cite[Proof of Proposition 3]{Ngo2013a}).

Furthermore, for $l \not \in  \Phi_j$ we have
\begin{equation}
\begin{split}
 \mathbb{E}_{\{\vect{h},\chi\}} \{ |\vect{g}_{jk}^{\Htran} \vect{h}_{jlm} |^2  \} & = d_j(\vect{z}_{lm}) \mathbb{E}_{\{\vect{h},\chi\}} \{ \| \vect{g}_{jk} \|^2\} \\ &=  \frac{1}{M-K_j} \frac{d_j(\vect{z}_{lm})}{d_j(\vect{z}_{jk})}  A_j,
\end{split}
\end{equation}
while for $l=j$ we obtain \eqref{eq:exp_gain3}, at the top of next page, 
\begin{figure*}[th]
\begin{equation}\label{eq:exp_gain3}
\begin{split}
 \mathbb{E}_{\{\vect{h},\chi\}} \{ |\vect{g}_{jk}^{\Htran} \vect{h}_{jjm} |^2  \} & =    \mathbb{E}_{\{\vect{h},\chi\}} \{ |\vect{g}_{jk}^{\Htran} \hat{\vect{h}}_{jjm} |^2  \}  +  \mathbb{E}_{\{\vect{h},\chi\}} \{ |\vect{g}_{jk}^{\Htran} \tilde{\vect{h}}_{jjm} |^2  \}   
= \frac{1}{M-K_j} \mathbb{E}_{\{\chi\}} \left\{  \frac{d_j(\vect{z}_{jm}) - \delta_{jjm}}{\delta_{jjk}}  \right\}  +  \begin{cases}
 1  & k=m \\
  0  & k \neq m
 \end{cases} \\
 & =   \begin{cases}
 1 + \frac{1}{M-K_j} (A_j - 1) & k=m \\
  \frac{1}{M-K_j} \left( \frac{d_j(\vect{z}_{jm})}{d_j(\vect{z}_{jk})} A_j - 
  \mathbb{E}_{\{\chi\}} \left\{  \frac{\delta_{jjm}}{\delta_{jjk}}  \right\} \right) & k \neq m
 \end{cases}
  \end{split}
\end{equation}
\noindent\rule{\textwidth}{0.4pt}
\begin{multline}\label{eq:SINR_lowerbound}
\mathrm{SINR}_{jk} \geq \\
 \frac{ 1 }{ 
\fracSum{l \in \Phi_j \setminus \{ j \}} \fracSumtwo{m=1}{K_l} 
 \frac{ (d_j(\vect{z}_{lm}) )^2}{  (d_l(\vect{z}_{lm}) )^2}  \frac{1}{| \Phi_j | B^\mathrm{cell}}  +
  \fracSum{l \in \mathcal{L}  } \fracSumtwo{m=1}{K_l} 
  \frac{1}{M-K_j} \frac{d_j(\vect{z}_{lm})}{d_l(\vect{z}_{lm})}  A_j - 
  \fracSum{l \in \Phi_j\setminus \{ j \}} \fracSumtwo{m=1}{K_l}   \frac{\left( \frac{d_j(\vect{z}_{lm})}{d_l(\vect{z}_{lm})} \right)^2 }{M-K_j}
 \frac{K_j}{| \Phi_j | B^\mathrm{cell}}   
  - \frac{K_j}{M-K_j}   +  \frac{\sigma^2}{\rho} \frac{A_j }{M-K_j} }
\end{multline}
\noindent\rule{\textwidth}{0.4pt}
\end{figure*}
where Jensen's inequality can be used to prove that 
\begin{equation} \label{eq:lower-bound-interference}
\mathbb{E}_{\{\chi\}} \left\{  \frac{\delta_{jjm}}{\delta_{jjk}}  \right\} \geq \frac{ \rho d_j(\vect{z}_{jm}) B A_j}{ \rho d_j(\vect{z}_{jk}) B A_j} = \frac{ d_j(\vect{z}_{jm}) }{ d_j(\vect{z}_{jk})}
\end{equation}
which leads to an upper bound on the interference term. In the same way, one can show that for $l  \in  \Phi_j \setminus \{ j \}$ we have
\begin{multline}
 \mathbb{E}_{\{\vect{h},\chi\}} \{ |\vect{g}_{jk}^{\Htran} \vect{h}_{jlm} |^2  \} \leq  \frac{ (d_j(\vect{z}_{lm}) )^2}{ d_j(\vect{z}_{jk}) d_l(\vect{z}_{lm})}  \frac{1}{| \Phi_j | B^\mathrm{cell}} \\ + \frac{1}{M-K_j} \frac{d_j(\vect{z}_{lm})}{d_j(\vect{z}_{jk})}  \left( A_j - \frac{K_j}{| \Phi_j | B^\mathrm{cell}}  \frac{d_j(\vect{z}_{lm})}{d_l(\vect{z}_{lm})} \right)
\end{multline}
where the inequality is due to \eqref{eq:lower-bound-interference}, $ \frac{1}{| \Phi_j | B^\mathrm{cell}} $ is the chance that a particular UE in another cell uses the same pilot sequence as UE $k$ in cell $j$, while $\frac{K_j}{| \Phi_j | B^\mathrm{cell}}$ is the chance that a particular UE in another cell uses any of the $K_j$ pilot sequences used in the cell $j$.
By plugging these expectations into \eqref{eq:SINR-value} and dividing all terms with $\frac{\rho}{d_j(\vect{z}_{jk})}$, we obtain the lower bound in \eqref{eq:SINR_lowerbound} at the top of next page. Next, we use Jensen's inequality in the same way as for MRC to move the expectation with respect to user positions to the denominator of the SINRs. The final expression in \eqref{eq:achievable-SINR-ZF2} follows from computing the expectations using \eqref{eq:position-averaging1}--\eqref{eq:position-averaging3}.

\bibliographystyle{IEEEtran}
\bibliography{IEEEabrv,refs}

\end{document}